%% file: main.tex
\documentclass[twocolumn]{aastex63}
\usepackage[utf8]{inputenc}
\usepackage{enumitem}
 
\usepackage{graphicx}
\usepackage{mathtools}
\usepackage{url}
\usepackage{hyperref}
\usepackage{natbib}
\usepackage{color}
\usepackage[normalem]{ulem}
\usepackage{academicons}
\usepackage{booktabs}
\usepackage{multirow}

\hypersetup{colorlinks=True,urlcolor=cyan,citecolor={blue},linkcolor={red}}

\begin{document}

\title[Source Confusion with Astro4cast]{Disentangling Multiple Stochastic Gravitational Wave Background Sources in PTA Datasets}

\input{authors_short}

\shortauthors{Kaiser et al.}

\correspondingauthor{Andrew Kaiser}
\email{andrew.kaiser@nanograv.org}

\begin{abstract}
    With strong evidence of a common-spectrum stochastic process in the most recent datasets from the NANOGrav Collaboration, the European Pulsar Timing Array (PTA), Parkes PTA, and the International PTA, it is crucial to assess the effects of the several astrophysical and cosmological sources that could contribute to the stochastic gravitational wave background (GWB).
    Using the same dataset creation and injection techniques as in \cite{Pol2021}, we assess the separability of multiple GWBs by creating single and multiple GWB source datasets.
    We search for these injected sources using Bayesian PTA analysis techniques to assess recovery and separability of multiple astrophysical and cosmological backgrounds.
    For a GWB due to supermassive black hole binaries and an underlying weaker background due to primordial gravitational waves with a GW energy density ratio of $\Omega_{\mathrm{PGW}}/\Omega_{\mathrm{SMBHB}} = 0.5$, the Bayes' factor for a second process exceeds unity at 17 years, and increases with additional data. 
    At 20 years of data, we are able to constrain the spectral index and amplitude of the weaker GWB at this density ratio to a fractional uncertainty of $64\%$ and $110\%$, respectively, using current PTA methods and techniques. 
    Using these methods and findings, we outline a basic protocol to search for multiple backgrounds in future PTA datasets.
\end{abstract}


\section{Introduction}
\label{sec:Intro}
\input{Section_Intro}

\section{Methods}
\label{sec:Methods}
\input{Section_Methods}

\section{Results}
\label{sec:Results}
\input{Section_Results}

\section{Discussion and Conclusions}
\label{sec:Discussion}
\input{Section_Discussion}

\newpage
\acknowledgements

This work has been carried out by the NANOGrav collaboration, which is part of the International Pulsar Timing Array. 
The NANOGrav project receives support from National Science Foundation (NSF) Physics Frontiers Center award number \#1430284 and \#2020265.
This work made use of the Super Computing System, Thorny Flat, at West Virginia University (WVU), which is funded in part by the  National Science Foundation (NSF) Major Research Instrumentation Program (MRI) Award \#1726534, and the Super Computing System Spruce Knob at WVU, which is funded in part by the National Science Foundation EPSCoR Research Infrastructure Improvement Cooperative Agreement \#1003907, the state of West Virginia (WVEPSCoR via the Higher Education Policy Commission) and WVU. SRT acknowledges support from NSF AST-200793, PHY-2020265, PHY-2146016, and a Vanderbilt University College of Arts \& Science Dean's Faculty Fellowship.

\section*{Software:} 
We use the software packages \texttt{Enterprise} \citep{Enterprise} and \texttt{enterprise\_extensions} \citep{enterpriseextensions} to setup our models, priors, and calculate the likelihoods. 
We use \texttt{PTMCMCSampler} \citep{PTMCMC} as the Markov Chain Monte Carlo sampler for our Bayesian analyses. 
We use \texttt{libstempo} \citep{libstempo} to inject the PTA noise parameters and GWB signals. 
We also use extensively, Matplotlib \citep{Matplotlib2007}, NumPy \citep{Numpy2020}, Python \citep{Python2007,Python2011}, and SciPy \citep{Scipy2020}.

\clearpage
\bibliographystyle{aasjournal}
%\bibliography{biblio.bib}
\input{biblio.tex}

\end{document}

%% file: authors_short.tex
% DO NOT EDIT THIS FILE. EDITS WILL BE OVERWRITTEN.
% AUTO-GENERATED WITH make-aastex62-author-list.py
% FROM ./author_list_12yr_data.txt, ./author_affil_and_orcid.txt, AND ./affil.txt

\author[0000-0002-3654-980X]{Andrew R. Kaiser}
\affiliation{Department of Physics and Astronomy, West Virginia University, P.O. Box 6315, Morgantown, WV 26506, USA}
\affiliation{Center for Gravitational Waves and Cosmology, West Virginia University, Chestnut Ridge Research Building, Morgantown, WV 26505, USA}
\author[0000-0002-8826-1285]{Nihan S. Pol}
\affiliation{Department of Physics and Astronomy, Vanderbilt University, 2301 Vanderbilt Place, Nashville, TN 37235, USA}
%\affiliation{Department of Physics and Astronomy, West Virginia University, P.O. Box 6315, Morgantown, WV 26506, USA}
%\affiliation{Center for Gravitational Waves and Cosmology, West Virginia University, Chestnut Ridge Research Building, Morgantown, WV 26505, USA}

\author[0000-0001-7697-7422]{Maura A. McLaughlin}
\affiliation{Department of Physics and Astronomy, West Virginia University, P.O. Box 6315, Morgantown, WV 26506, USA}
\affiliation{Center for Gravitational Waves and Cosmology, West Virginia University, Chestnut Ridge Research Building, Morgantown, WV 26505, USA}

\author[0000-0002-3118-5963]{Siyuan Chen}
\affiliation{Kavli Institute for Astronomy and Astrophysics, Peking University, Beijing 100871, China}

\author[0000-0003-2742-3321]{Jeffrey S. Hazboun}
%\altaffiliation{NANOGrav Physics Frontiers Center Postdoctoral Fellow}
\affiliation{University of Washington Bothell, 18115 Campus Way NE, Bothell, WA 98011, USA}

\author[0000-0002-6625-6450]{Luke Zoltan Kelley}
\affiliation{Center for Interdisciplinary Exploration and Research in Astrophysics (CIERA), Northwestern University, Evanston, IL 60208}

\author[0000-0003-1407-6607]{Joseph Simon}
\affiliation{Department of Astrophysical and Planetary Sciences, University of Colorado, Boulder, CO 80309, USA}

\author[0000-0003-0264-1453]{Stephen R. Taylor}
\affiliation{Department of Physics and Astronomy, Vanderbilt University, 2301 Vanderbilt Place, Nashville, TN 37235, USA}

\author[0000-0003-4700-9072]{Sarah J. Vigeland}
\affiliation{Center for Gravitation, Cosmology and Astrophysics, Department of Physics, University of Wisconsin-Milwaukee,\\ P.O. Box 413, Milwaukee, WI 53201, USA}
\author[0000-0002-6020-9274]{Caitlin A. Witt}

\affiliation{Department of Physics and Astronomy, West Virginia University, P.O. Box 6315, Morgantown, WV 26506, USA}
\affiliation{Center for Gravitational Waves and Cosmology, West Virginia University, Chestnut Ridge Research Building, Morgantown, WV 26505, USA}
%\author[0000-0003-1096-4156]{William Lamb}
%\affiliation{Department of Physics and Astronomy, Vanderbilt University, 2301 Vanderbilt Place, Nashville, TN 37235, USA}
%
%
%\author{ The NANOGrav Collaboration}

%% file: Section_Intro.tex
Predicted sources of low-frequency gravitational wave backgrounds (GWBs) range from the most massive binary systems in the universe to the quantum fluctuations stretched to super-horizon size during the epoch of cosmic inflation in the early universe.
With the recent observation of a common-spectrum process in several pulsar timing array (PTA) datasets, many potential astrophysical and cosmological sources have been proposed (see, e.g.,  \cite{Vaskonen2021,DeLuca2021,Ellis2021,Blasi2021,Bhattacharya2021}). 
The primary expected source of a low-frequency gravitational wave background (GWB) is a population of supermassive black hole binaries (SMBHBs; \cite{Rajagopal1995,Jaffe2003,Sesana2004,BurkeSpolaor2019}).
In addition to SMBHBs, more exotic sources could contribute to a common-spectrum process including: cosmic strings \citep{Kibble1976,Vilenkin1981,Vilenkin1985,Vilenkin2000,Siemens2007,Olmez2010,BlancoPillado2018}, primordial gravitational waves \citep{Grishchuk1976,Grishchuk1977,Starobinsky1980,Linde1982,Fabbri1983,Grishchuk2005,Lasky2016}, and phase transitions in the early universe \citep{Winicour1973,Hogan1986,Deryagin1986,Caprini2010,Kobakhidze2017,12p5yr_pt}.

Spectra of stochastic GWBs are often characterized by the relative GW energy density per logarithmic frequency to the critical density required to close the universe,
\begin{equation}
    \Omega_{\mathrm{gw}}(f) \equiv \frac{1}{\rho_{\mathrm{c}}}\frac{\mathrm{d}\rho_{\mathrm{gw}}(f)}{\mathrm{d}~\mathrm{ln}f}
\end{equation}
where $f$ is the observed frequency, $\rho_{\mathrm{c}} \equiv 3c^{2}H_{0}^{2}/8\pi G$ is the critical density, $H_{0} = 100h~ \mathrm{km}~\mathrm{s}^{-1}~\mathrm{Mpc}^{-1}$ is the Hubble expansion rate, and we use the same assumption as \cite{Lasky2016} of $h=0.67$ as the dimensionless Hubble parameter \citep{Planck2020}.
Our conclusions do not depend on the value of the dimensionless Hubble parameter and thus should be independent of the tension associated with $h$.

The GW energy density can also be expressed as the strain power spectral density
\begin{equation}
    S_{h}(f) = \frac{3H_{0}^{2}}{2\pi^{2}}\frac{\Omega_{\mathrm{gw}}(f)}{f^{3}} ~,
\end{equation}
or characteristic strain by
\begin{equation}
    h_{\mathrm{c}}(f) =\sqrt{fS_{h}(f)} ~.
\end{equation}
For a simple power-law spectrum, the form of the dimensionless energy-density spectrum can be simplified to
\begin{equation}
    \Omega_{\mathrm{gw}}(f) = \Omega_{\beta}\left(\frac{f}{f_{\mathrm{ref}}}\right)^{\beta} ~,
\end{equation}
where $\beta$ is the spectral index, $\Omega_{\beta}$ is the amplitude of the energy density, and $f_{\mathrm{ref}}$ is an arbitrary reference frequency, which PTAs often assume to be $f_\mathrm{ref}=1/\mathrm{yr}$.
Thus the characteristic strain becomes the familiar
\begin{equation}
    h_{\mathrm{c}}(f)=A_{\alpha}\left(\frac{f}{f_{\mathrm{ref}}}\right)^{\alpha} ~,
\end{equation}
where 
\begin{equation}
    \Omega_{\beta} = \frac{2\pi^{2}}{3H_{0}^{2}}f_{\mathrm{ref}}^{2}A_{\alpha}^{2}
\end{equation}
relates the strain spectral index $\alpha$ to $\beta$ via $\beta = 2\alpha + 2$ and the strain amplitude, $A_{\alpha}$ to $\Omega_{\beta}$ \citep{Thrane2013}. 

For working with PTA data, the power-law form is commonly transformed again in terms of the timing-residual cross-power spectral density
\begin{equation}
    S_{\mathrm{ab}}(f) = \Gamma_{\mathrm{ab}}~\frac{A_{\alpha}^{2}}{12\pi^{2}}\left(\frac{f}{f_{\mathrm{ref}}}\right)^{-\gamma}~f_{\mathrm{ref}}^{-3} ~,
\end{equation}
where $\gamma\equiv3-2\alpha$, and $\Gamma_{\mathrm{ab}}$ is the overlap reduction function that represents the expectation value of the inter-pulsar spatial cross-correlation.
For an isotropic GWB, the overlap reduction function takes the form of the Hellings-Downs correlation \citep{Hellings1983}.
We often express this in residual space, and thus it becomes the GWB residual delay
\begin{equation}
    \label{eq:GWB_res_delay}
    \rho(f) = \sqrt{\frac{S_{\mathrm{ab}}(f)}{T_{\mathrm{obs}}}} ~,
\end{equation}
where $T_{\mathrm{obs}}$ is the longest observing timescale in seconds.

We do not expect the cross-correlations to provide much additional evidence until baselines longer than 20 years, so the auto-correlation terms provide the bulk of the information at the PTA timescales examined here \citep{Romano2021,Pol2021}.
Thus in this work, we restrict our injections to the simpler case of only auto-correlations for each background type and do not take into account the cross-correlations due to the above considerations and significant computational costs.
In all of this work we assume that there is a detectable GW signal, and are only concerned with its origin.

\subsection{Gravitational Wave Background Sources in the PTA Band}
\label{subsec:GWBSourcesPTABand}
    The GWB for each predicted source in the PTA frequency band has a characteristic predicted spectral shape.
    In this work, we focus on two main sources: SMBHBs, because they are predicted to be the primary source of a low-frequency GWB, and primordial gravitational waves (PGWs), for their diversity of spectral indices both steeper and shallower than a GWB from SMBHBs.
        \subsubsection{Supermassive Black Hole Binaries}
        \label{subsubsec:SMBHBs}          
            SMBHBs are predicted to be the most plausible signal in the PTA frequency regime (for our work between  $\sim$ 2 nHz -- 80 nHz) \citep{Rajagopal1995,Jaffe2003,Sesana2004,BurkeSpolaor2019}.
            SMBHBs in their early inspiral, where their component black holes' orbits slowly evolve at large separations, are potentially individually resolvable with PTAs, but are likely not yet detectable at current PTA sensitivities \citep{Rosado2015,Mingarelli2017,Kelley2018}. 

            In a universe filled with many individual binaries (in which we expect to find ourselves), they each contribute stochastically to a superposition of all of their signals \citep{Rajagopal1995,Jaffe2003,Sesana2004,BurkeSpolaor2019}.
            While the exact shape of the spectrum from SMBHBs is dependent on many factors (gas and stellar environment, eccentricity, etc.) \citep{Kelley2017,Taylor2017,Chen2019}
            we assume a GW background from SMBHBs is characterized by a continuous distribution of circular binaries evolving purely due to GW emission.
            These assumptions lead to a simple power-law spectrum over all frequencies in the PTA band with $\alpha=-2/3$ ($\gamma=13/3$) \citep{Rajagopal1995,Phinney2001,Jaffe2003,Sesana2008}.
            Because GW radiation from the stochastic SMBHB background is expected to be the brightest GWB source in the PTA band, in this work we treat it as a foreground when examining the possibility of multiple backgrounds.
        
        \subsubsection{Primordial Gravitational Waves}
        \label{subsubsec:PGWs}
            PGWs originate from quantum fluctuations stretched to super-horizon size during the epoch of cosmic inflation in the early universe.
            The first attempts at indirectly detecting PGWs were made by searching for a characteristic polarization in the cosmic microwave background (see \cite{Kamionkowski2016} for a review).
            Due to the early times at which these PGWs are generated, their spectrum is tied to the particular model of inflation used and the equation of state of the early universe immediately after inflation by
            \begin{equation}
                \alpha = \frac{n_{t}}{2} - \frac{2}{3w+1} ,
            \end{equation}
            where $n_{t}$ is the tensor index of the primordial power spectrum and $w$ is the equation-of-state (EOS) parameter of the early universe \citep{9yr_gwb,Lasky2016}.
            
            There are a wide range of combinations of tensor index and EOS that produce spectral indices around that of an SMBHB GWB.
            Since this work is restricted to power-law spectra for GWBs in the PTA frequency band, the particular values of the tensor index and EOS are not evaluated here.
            We consider two plausible spectral indices resulting from PGWs: $\alpha=-1$ ($\gamma=5$) and $\alpha=-1/2$ ($\gamma=4$).
            Using these two trial spectral indices has the added benefit of allowing the examination of spectra both steeper and shallower than the predicted spectrum from SMBHBs ($\alpha=-2/3$, $\gamma=13/3$).
            Upper limits have been set on each of these sources of PGWs in PTAs, however each assume PGWs are the only source when setting limits.
            \cite{11yr_gwb} sets an upper limit using the NANOGrav 11-year dataset of
            \begin{equation}
                \Omega_{\mathrm{gw}}(f)h^{2} \leq 3.4(1)\times10^{-10}
            \end{equation}
            where $h$ is the dimensionless Hubble parameter, for PGW GWB with spectral index $\alpha=-1$.
            \cite{Lasky2016} sets a tighter constraint of 
            \begin{equation}
                \Omega_{\mathrm{gw}}(f)h^{2} \leq 1.0\times10^{-10}
            \end{equation}
            after combining multiple limits from CMB, LIGO/Virgo, and PTA GW experiments. 
            Because this work only looks at 20 years of simulated data, our PGW injections are all above $\Omega_{\mathrm{gw}}(f)h^{2} \geq 2.5\times10^{-10}$ to be able to make reasonable comparisons to the injected SMBHB of $\Omega_{\mathrm{gw}}(f)h^{2} \sim 2.5\times10^{-9}$.
            Future work with extended baseline datasets will be able to look at realistic amplitudes below current limits.

%% file: Section_Methods.tex
\subsection{Simulated Data}
\label{subsec:SimulatedData}
    In this paper, we use the same methods of simulating realistic PTA data as \cite{Pol2021}, which is rooted in the characteristics of the NANOGrav 12.5-yr dataset \citep{12p5yr_gwb}.
    The times-of-arrival (TOAs) in the simulated dataset are the same as those in the real dataset.
    The radiometer uncertainties and pulse-phase jitter noise are determined using the maximum likelihood pulsar noise estimates made as part of the 12.5-yr analysis. 
    Each pulsar also contains intrinsic red noise (RN), which was derived by modeling the intrinsic RN alongside a GWB-like $\alpha = -2/3$ process in order to filter out any covariance with the common process detected in the real 12.5-yr dataset.
    
    The dataset is then extended into the future up to a baseline of 20 years by drawing TOAs from distributions of the observational cadence and TOA uncertainties from the last year of the 12.5-yr dataset. We retain all the 45 pulsars used in the 12.5-yr analysis, and do not add new pulsars in these simulations. As described in \citet{Pol2021}, this implies that growth rate of the detection statistics in this work will be a conservative estimate, since the addition of pulsars is known to improve the ability of PTAs to detect and characterize the GWB \citep{Siemens2013}.
    
    After injecting the intrinsic pulsar RN, we inject 50 realizations of multiple stochastic GWBs in this dataset to forecast the detection, characterization, and separability of these GWB signals. 
    When injecting multiple GWBs, we inject two separate stochastic signals. 
    We use \texttt{libstempo}, a Python wrapper for \texttt{TEMPO2}, to inject each GWB \citep{TEMPO2I,TEMPO2II,TEMPO2III,libstempo}.
    Each GWB is injected over the duration of the entire dataset at frequencies from $1/(10~T_{\mathrm{obs}})$, where $T_{\mathrm{obs}}$ in this case is taken to be our longest baseline of 20 years, to the Nyquist frequency (half of the sampling rate), which we assume to be half of the observing cadence for a cadence of once per two weeks, with linear spacing of $1/(10~T_{\mathrm{obs}})$.
    The amplitude and spectral index of each GWB is then constructed in residual space as a pure power-law over these frequencies.
    Once in residual space, the power-law is used to scale the Hellings-Downs overlap reduction function that is constructed based on the position of the pulsars with random zero-mean, unit variance Gaussian shifts in the frequency series.
    The resulting residuals in the frequency domain are then transformed via an inverse Fourier Transform to the TOA space and injected into each pulsar's TOAs over its corresponding observed timespan.

    Finally, in order to mimic the PTA analysis pipeline, we then refit the timing models of each pulsar using \texttt{TEMPO2} to extract a new parameter file that is then used in the analyses. 
    
\subsection{Analysis Methods}
\label{subsec:AnalysisMethods}
    In this work, we use similar techniques to the standard analysis methods for PTA GW detection (e.g. \citet{9yr_gwb,11yr_gwb,12p5yr_gwb}). 
    We restrict our searches to a time-correlated process that is common to all pulsars, but ignoring spatial correlations. 
    This drastically speeds up the analysis time, and does not affect the test of our spectral analysis's ability to differentiate between processes with different spectral indices. 
    Searching for spatial correlations would allow us to isolate the signal further, hence making these simulations a conservative assessment of this search when correlations are detectable.

    Each type of analysis and their combinations include noise for each pulsar.
    The white noise \citep[EFAC, EQUAD, ECORR][]{5yr_bwm, 9yr_gwb}, is applied to the TOA uncertainties when producing the simulated datasets and thus it is already incorporated in the analysis.
    Thus only the intrinsic RN (e.g. the spin  noise) of each pulsar is necessary to model in addition to the GWB.
    We model the pulsar RN as a single power-law over 30 Fourier-basis frequencies linearly spaced starting at $1/T_\mathrm{obs}$ and separated by $1/T_\mathrm{obs}$ with a varying spectral index and amplitude.
    Since the injected GWBs are more densely spaced ($1/(10T_\mathrm{obs})$) and shifted, we are searching over different frequencies than we injected.
    In order to improve the sampling and convergence of the $90$ pulsar RN parameters, we employ jump proposals from empirical distributions like those in \cite{11yr_cw}.
    To do so, we first run the free spectrum base model (see section \ref{par:FreeSpec}) with all pulsars until their RN parameters are reasonably converged (auto-correlation lengths $<$ 1000, and a Gelman-Rubin split R hat statistic $<$ 1.1 \citep{Vehtari2019}), then construct 2-dimensional empirical distributions for the amplitude and spectral index of each pulsar.
    During subsequent runs, we include jump proposals that draw from these 2-D empirical distributions to reduce the number of samples without affecting the probability of future parameter value draws.
    
    \subsubsection{Base Models}
\begin{figure}[!htbp]
            \includegraphics[width=\columnwidth]{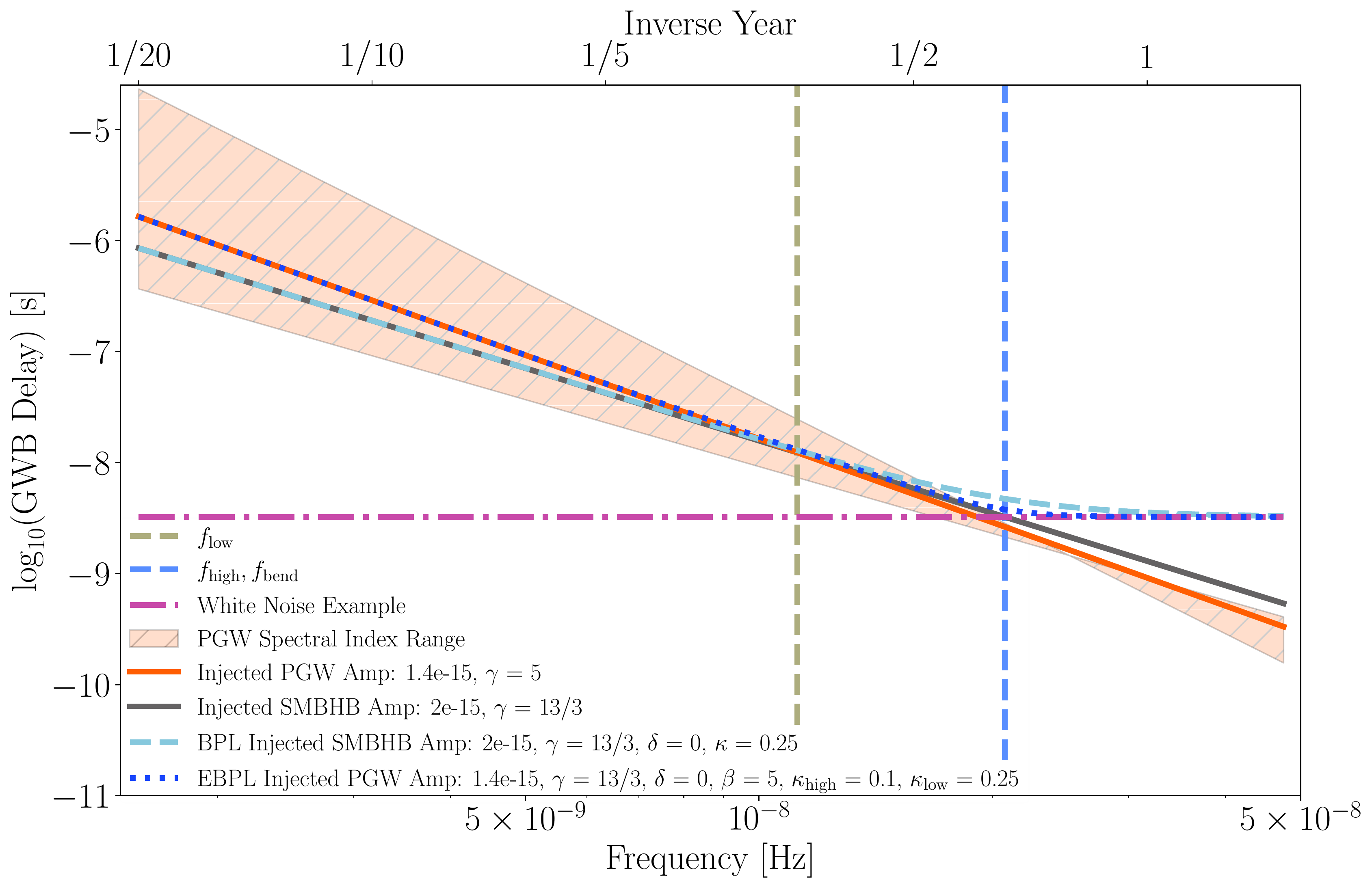}
            \caption{Examples of various base models used in this analysis.
            PLs for the injected SMBHB (solid grey line), example WN (dark pink dashed-dotted line), and PGW (solid orange line) with the range in spectral indices for the three early universe models covered in the orange region.
            A example BPL encapsulating the combination of SMBHB and WN PLs (cyan dashed line) with the bend frequency marked with the dashed light blue line, and an example EPBL combines the PGW, SMBHB, and WN PLs (dotted dark blue line) with the low and high frequency breaks marked in dashed olive green and light blue, respectively.
            Amplitudes are quoted at $f_\mathrm{ref}=1/\mathrm{yr}$.
            }
            \label{fig:PL_examples}
        \end{figure}

        \input{Tables/Base_Model_Types}
        Throughout this work we use five types of models for the GWB spectrum summarized in table \ref{tab:base_model_types} with examples plotted in figure \ref{fig:PL_examples}:
        \paragraph{Power-law}
        We use two subtypes of a power-law (PL) with the form in Section \ref{sec:Intro}.
        The fixed PL, a PL with freely varying amplitude but fixed spectral index, allows us to extract the power at a particular spectral index.
        This is particularly useful when we assume that we have detected a foreground signal and are seeking an additional PL at a different spectral index.
        The second sub-type is the free PL, with freely varying amplitude and spectral index.
        In the same example of a foreground and a background, we can use a varying spectral index to determine the amplitude and spectrum for the unknown background.
        This free PL model is especially useful at determining if there is evidence for extra power not associated with the dominant GWB source.
        In addition to restricting or freeing the spectral index, we can use models that employ frequency restrictions to better separate excess white noise from the power-law processes much like the technique used in \cite{12p5yr_gwb}.
        For this particular study we use 30 frequency components.
        As an additional test, we performed the same analyses on five frequencies, but found minimal difference in the results.
    
        \paragraph{Free Spectrum}
        \label{par:FreeSpec}
        The free spectral model allows us to analyze the power at each of the sine-cosine pairs in the RN Fourier basis independently.
        This method is particularly useful as it is agnostic of the source model; thus it does not require the prescription of a particular spectral model.
        It also can be used to assess the number of frequencies at which each power-law or noise is dominant, if there are frequencies with excess power, and how well other multi-frequency models capture the overall shape of each pulsar's spectrum. 
        
        \paragraph{Broken Power-law}
        The broken power-law (BPL) model, introduced by \cite{Sampson2015}, allows for a smooth transition between two PLs of the form 
        \begin{equation}
            \label{eq:brokenplaw}
            S(f) = \frac{A_{\gamma}^{2}}{12\pi^{2}}
            \left(\frac{f}{f_{\mathrm{ref}}}
            \right)^{-\gamma}
            \left[1+
            \left(\frac{f}{f_{\mathrm{bend}}}
            \right)^{1/\kappa}
            \right]^{\kappa(\gamma-\delta)}f_{\mathrm{ref}}^{-3} ,
        \end{equation}
        where $\gamma$ is the spectral index of the power-law at frequencies lower than $f_{\mathrm{bend}}$ and $\delta$ is the slope at higher frequencies, $\kappa$ captures the smoothness of the transition between the two, $A_{\gamma}$ is the amplitude of the PL governed by the $\gamma$ spectral index, and $f_{\mathrm{ref}}$ is an arbitrary reference frequency, which we assume to be $f_\mathrm{ref}=1/\mathrm{yr}$.
        In a dataset where the GW spectrum is composed of several processes, a BPL is ideal to determine the transition frequencies between two processes.
        In general, the second, high frequency PL is the ever-present white noise in the detector. 
        We use the BPL much like \cite{12p5yr_gwb} to determine the optimal number of frequencies at which a low frequency PL dominates over the white noise that is omnipresent at higher frequencies.
        
        \paragraph{Extra-broken Power-law}
        This new extra-broken power-law (EBPL) model allows for a smooth transition between three power-laws of the form 
        \begin{widetext}
            \begin{equation}
                \label{eq:extrabrokenplaw}
                S(f) = \frac{A_{\beta}^{2}}{12\pi^{2}}
                \left(\frac{f}{f_{\mathrm{ref}}}\right)^{-\beta}
                \left\{
                \left[ 1+ 
                \left(\frac{f}{f_{\mathrm{low}}}
                \right)^{-\kappa_{\mathrm{low}}} 
                \left(
                \left[1+
                \left(\frac{f}{f_{\mathrm{high}}}
                \right)^{-\kappa_{\mathrm{high}}}
                \right]^{\frac{\kappa_{\mathrm{high}}
                \left(\gamma-\delta
                \right)}{\kappa_{\mathrm{low}}
                \left(\beta-\gamma
                \right)}}
                \right)
                \right]^{\kappa_{\mathrm{low}}
                \left(\beta-\gamma
                \right)/2}
                \right\}
                f_{\mathrm{ref}}^{-3} ,
            \end{equation}
        \end{widetext}
        where $\beta$ is the spectral index of the PL at frequencies lower than $f_{\mathrm{low}}$, $\gamma$ is the spectral index at frequencies between  $f_{\mathrm{low}}$ and $f_{\mathrm{high}}$, and $\delta$ is the slope at higher frequencies, $\kappa_{\mathrm{low}}$ and $\kappa_{\mathrm{high}}$ capture the smoothness of the transition between the at $f_{\mathrm{low}}$ and $f_{\mathrm{high}}$, respectively, $A_{\beta}$ is the amplitude of the PL governed by the $\beta$ spectral index, and $f_{\mathrm{ref}}$ is an arbitrary reference frequency, which we assume to be $f_\mathrm{ref}=1/\mathrm{yr}$.
        We developed this broken power-law model with an extra break to better differentiate between three power-laws.
        This will allow us to probe whether we can identify the PGW-dominated, SMBBH-dominated, and WN-dominated regimes.
        It behaves exactly like the BPL model, but with an extra transition at the lowest frequencies to a steeper PL. 
    
    \subsubsection{Model Comparisons}
        \input{Tables/Model_Comparison_Types}
        We can directly compare two different models for the GWB by fitting for two models with common parameters simultaneously using product space sampling \citep{Carlin1995,Godsill2001,Hee2016,11yr_gwb}. 
        We use the same methodology for product space sampling (hereafter called ``hypermodels") as in \cite{Taylor2020}.
        The crucial model mixing parameter is a hyper parameter that independently samples between the two models and builds up samples in both. 
        The ratio of the samples between the two models directly corresponds to the Bayes' factor (BF) and indicates whether one model is preferred over the other. 
        It also saves computation power by burning in common parameters (e.g., the intrinsic pulsar RN and the GW amplitude) while switching between unique parameters. 
        For example, we examined the preference of a single 30 frequency fixed PL corresponding to the SMBHB index versus the same 30 frequency fixed PL with an additional 30 frequency free PL model (hypermodel A in table \ref{tab:model_comp_types}). 
        The latter model only added two parameters (and the parameter to sample between models) and can demonstrate the preference of the presence of an additional common RN process in the data.
        
        We use four types of hypermodels for the GWB spectrum outlined in table \ref{tab:model_comp_types}: 
        \paragraph{Hypermodel A}
        \label{par:fixedvsfree}
            One of the first tests done on new datasets to investigate the significance of a particular single spectral index PL at a relatively low computational cost is the fixed PL.
            This is distinctly useful in this work since we assume there is a foreground with a spectral index of $\alpha=-2/3$ ($\gamma=13/3$).
            In order to test the hypothesis of a background spectrum, we set up an additional model in a hypermodel framework to allow for the inclusion of a free PL model on top of the fixed PL model.
            If the hypermodel shows significance (i.~e. a high BF) for including the additional free PL, particularly if the spectral characteristics are well constrained, we conclude there is excess power at lower frequencies that should be looked at more carefully to conclude if it is indeed a physical signal.
        
        \paragraph{Hypermodel B}
            To discover the amount of power associated with a particular spectral index, we follow-up the hypermodel A with one that is more restrictive.
            We wish to find if two fixed power-laws  are preferred to a single fixed PL with our hypermodel B.
            This is a direct follow-up from our previous model since power in the free PL can soak up power from the fixed PL, or vice-versa.
            This search allows us to pinpoint the amount of power at a particular index to help mitigate cross-contamination of signal.
            It is important to note that this test should be secondary to a free PL as fixed PLs make a strong assumption of a particular index being present in the data.
            
        \paragraph{Hypermodel C}
            In the event of finding significance of an additional process with either a hypermodel A or B, one should be sure to examine that the background ``signal" picked up by the fixed PLs is actually favored.
            Hypermodel C is a further combination of fixed and free PLs where the selection is between just a free PL and the combination of fixed and free PLs.
            This setup allows one to be sure the fixed PL is assumed to be at the correct spectral index by testing the evidence for the inclusion of a fixed PL on top of just a free PL.
            
        \paragraph{Hypermodel D}
            For this hypermodel setup we compare a 30-frequency BPL versus a single free PL.
            This model comparison gives us information about whether the spectrum has a distinct break frequency.
            \cite{12p5yr_gwb} used the BPL model to provide support for a steep red process present at the lowest frequencies, but still account for the flatter spectrum from the superposition of various noise processes at higher frequencies. 
            We employ the same strategy in the hypermodel framework to determine both whether the spectrum contains a distinct break (i.~e. preferring a BPL to a simple free PL) and around which frequencies the break occurs, if a break is preferred, to inform our other models.
            
        \paragraph{Hypermodel E}
            Finally, we construct hypermodels to compare the BFs of two break frequencies in the data (i.~e. three distinct PLs, the EBPL) to a free PL. 
            This introduces another degree of freedom in characterizing the spectral properties of the data.
            If there is significant evidence for three separate power-laws; one at high frequencies for the white noise, another at mid-range frequencies marking the foreground spectral process, and one at the lowest frequencies for a steeper background process, then this particular model setup allows us to compare to both the models used in usual detection techniques.
            While it may be some time before current PTAs have the sensitivity to discriminate three backgrounds, this new model is applicable beyond this study in future searches.
            \cite{12p5yr_gwb} fit a five-frequency power-law model and found a common spectral process with  a marginally steeper index at the one-sigma level than predicted from a GWB made up of only SMBHBs. 
            By implementing the EBPL model, one could determine whether there is evidence for multiple low-frequency power-laws present in the data, in particular, one with a steeper spectral index at the lowest frequencies.

%% file: Tables/Base_Model_Types.tex
\begin{table*}[!htbp]
    \caption{Base Model Types}
    \scalebox{0.9}{
    \noindent\makebox[\textwidth]{
        \begin{tabular}{@{} c|ccccc @{}}
            \hline\hline
               & Extra-broken Power-law & Broken Power-law & Free Power-law & Fixed Power-law & Free Spectrum \\
            \hline
                Intrinsic Pulsar Noise Parameters &  &  &  &  & \\
            \hline
                $A_{\mathrm{red}}$ & \checkmark & \checkmark & \checkmark & \checkmark & \checkmark  \\
                $\gamma_{\mathrm{red}}$ & \checkmark & \checkmark & \checkmark & \checkmark & \checkmark \\
            \hline
                GWB Power-law Parameters &  &  &  &  & \\
            \hline
                $\rho(f_{i})$ & - & - & - & - & \checkmark \\
                $A$ & \checkmark & \checkmark & \checkmark & \checkmark & - \\
                $\gamma$ & \checkmark & \checkmark & \checkmark & - & - \\
                $\delta$ & \checkmark & \checkmark & - & - & - \\
                $\kappa,(\kappa_{\mathrm{low}})$ & \checkmark & \checkmark & - & - & - \\
                $f_{\mathrm{bend}},(f_{\mathrm{high}})$ & \checkmark & \checkmark & - & - & - \\
                $\kappa_{\mathrm{high}}$ & \checkmark & - & - & - & - \\
                $f_{\mathrm{low}}$ & \checkmark & - & - & - & - \\
    
        \end{tabular}
    }}
    \tablecomments{Different base model types performed in our analyses.}
    \label{tab:base_model_types}
\end{table*}

%% file: Tables/Model_Comparison_Types.tex
\begin{table*}[!htbp]
    \caption{Model Selection Types}
    \scalebox{0.9}{
    \noindent\makebox[\textwidth]{
        %\centering
        \begin{tabular}{@{} c|ccccc @{}}
            \hline\hline
               & Hypermodel A & Hypermodel B & Hypermodel C & Hypermodel D & Hypermodel E\\
            \hline
                Signal Model 1 & & & & \\
            \hline
                First Base Model & Fixed PL & Fixed PL & Free PL & Free PL & Free PL \\
                $A$ & Log\text{--}uniform [\text{--}18,\text{--}11] & Log\text{--}uniform [\text{--}18,\text{--}11] & Log\text{--}uniform [\text{--}18,\text{--}11] & Log\text{--}uniform [\text{--}18,\text{--}11] & Log\text{--}uniform [\text{--}18,\text{--}11] \\
                $\gamma$ & 13/3 & 13/3 & Uniform [0,8] & Uniform [0,8] & Uniform [0,8] \\
            \hline
                Signal Model 2 & & & & & \\
            \hline
            	First Base Model & Fixed PL & Fixed PL & Fixed PL & BPL & EBPL \\
            	$A$ & Log\text{--}uniform [\text{--}18,\text{--}11] & Log\text{--}uniform [\text{--}18,\text{--}11] & Log\text{--}uniform [\text{--}18,\text{--}11] & Log\text{--}uniform [\text{--}18,\text{--}11] & Log\text{--}uniform [\text{--}18,\text{--}11] \\
            	$\gamma$ & 13/3 & 13/3 & 13/3 & Uniform [0,8] & Uniform [0,8] \\
            	$\delta$ & \text{--} & \text{--} & \text{--} & Uniform [0,8] & Uniform [0,8] \\
                $\kappa,(\kappa_{\mathrm{low}})$ & \text{--} & \text{--} & \text{--} & Uniform [0.01,0.5] & Uniform [0.01,0.5] \\
                $f_{\mathrm{bend}},(f_{\mathrm{high}})$ & \text{--} & \text{--} & \text{--} & Log\text{--}uniform [\text{--}8,\text{--}7] & Log\text{--}uniform [\text{--}8,\text{--}7] \\
                $\kappa_{\mathrm{high}}$ & \text{--} & \text{--} & \text{--} & \text{--} & Uniform [0.01,0.5] \\
                $f_{\mathrm{low}}$ & \text{--} & \text{--} & \text{--} & \text{--} & Log\text{--}uniform [\text{--}10,\text{--}8] \\
                Second Base Model & Free PL & Fixed PL & Free PL & \text{--} & \text{--} \\
                $A$ & Log\text{--}uniform [\text{--}18,\text{--}11] & Log\text{--}uniform [\text{--}18,\text{--}11] & Log\text{--}uniform [\text{--}18,\text{--}11] & \text{--} & \text{--} \\
                $\gamma$ & Uniform [0,8] & $\gamma_{\mathrm{PGW}}~(4,5)$  & Uniform [0,8] & \text{--} & \text{--} \\
        \end{tabular}
    }}
    \tablecomments{Different model selection types performed in our analyses or recommendations. All power\text{--}law types include 30 frequencies to model the power\text{--}law.}
    \label{tab:model_comp_types}
\end{table*}

%% file: Section_Results.tex
Here we discuss the results of our analyses on the injected datasets.
In section \S\ref{subsec:MultipleSources} we first analyze injections with two GWB signals.
We use a hypermodel A to compute BFs and fractional uncertainties and then perform hypermodel D and E analyses to determine whether there is evidence for break frequencies in the data. 
The analyses in this section allows us to asses0s directly the separability of two GWB signals.
In section \S\ref{subsec:SingleGWBSources} we analyze a single injection of a PGW GWB to determine if our analyses are robust enough to assure us we will not misconstrue a single signal with our multiple signal model framework.

\subsection{Multiple Sources}
    \label{subsec:MultipleSources}
    Here we examine the injected backgrounds of an SMBHB GWB with a spectral index of $\alpha=-2/3$ and a PGW GWB with a spectral index of $\alpha=-1$.
    We inject the SMBHB GWB at an amplitude of $2\times10^{-15}$ at a frequency of $f_\mathrm{ref}=1/\mathrm{yr}$, which is consistent with the amplitude measured for the common-spectrum stochastic process by \cite{12p5yr_gwb}.
    For the PGW GWB we inject a series of amplitudes: $A_{\mathrm{yr}}=1.4\times10^{-15}$, $1.2\times10^{-15}$, $1\times10^{-15}$, $8.4\times10^{-16}$, and $6.3\times10^{-16}$, corresponding to GW density ratios of $\Omega_{\mathrm{PGW}}/\Omega_{\mathrm{SMBHB}} = 0.5$, $0.375$, $0.25$, $0.125$, and $0.1$, respectively at a reference frequency of $f_\mathrm{ref}=1/\mathrm{yr}$. 

    \begin{figure}[!htbp]
        \includegraphics[width=\columnwidth]{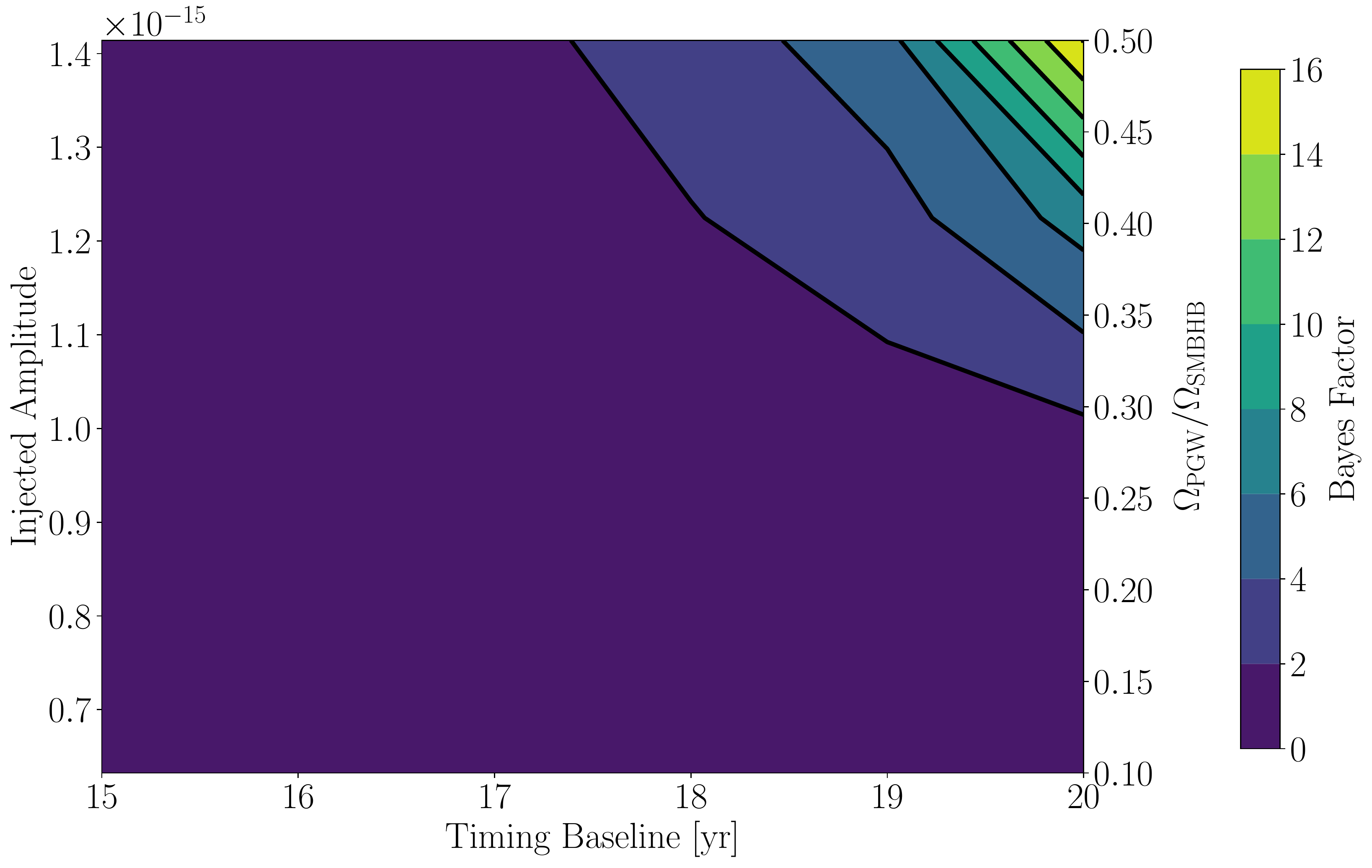}
        \caption{Median Bayes' factors for 50 realizations of an additional process in addition to a SMBHB GWB PL with respect to injected strain amplitudes at $f_\mathrm{ref}=1/\mathrm{yr}$ (left axis) and GW density fraction (right axis) versus the timing baseline of the PTA.
        }
        \label{fig:contour30fixedand30free}
    \end{figure}
    
    Figure \ref{fig:contour30fixedand30free} shows the Bayes' factors as a function of PTA timing baseline and injected PGW GWB amplitude for the hypermodel A.
    As the timing baseline increases, irrespective of the amplitude injected, the evidence for an extra process increases. 
    After around the 17 year mark, the greater the injected PGW GWB amplitude with respect to the SMBHB GWB amplitude, the greater evidence there is for an extra process.
    While the median BFs for all 50 realizations at all baselines remain low from a Bayesian perspective, in $22\%$ of realizations of the 20 year baseline for the $\Omega_{\mathrm{PGW}}/\Omega_{\mathrm{SMBHB}} = 0.5$ injection the BF for an additional process is greater than $1000$. 
    
    \begin{figure}[!htbp]
        \includegraphics[width=\columnwidth]{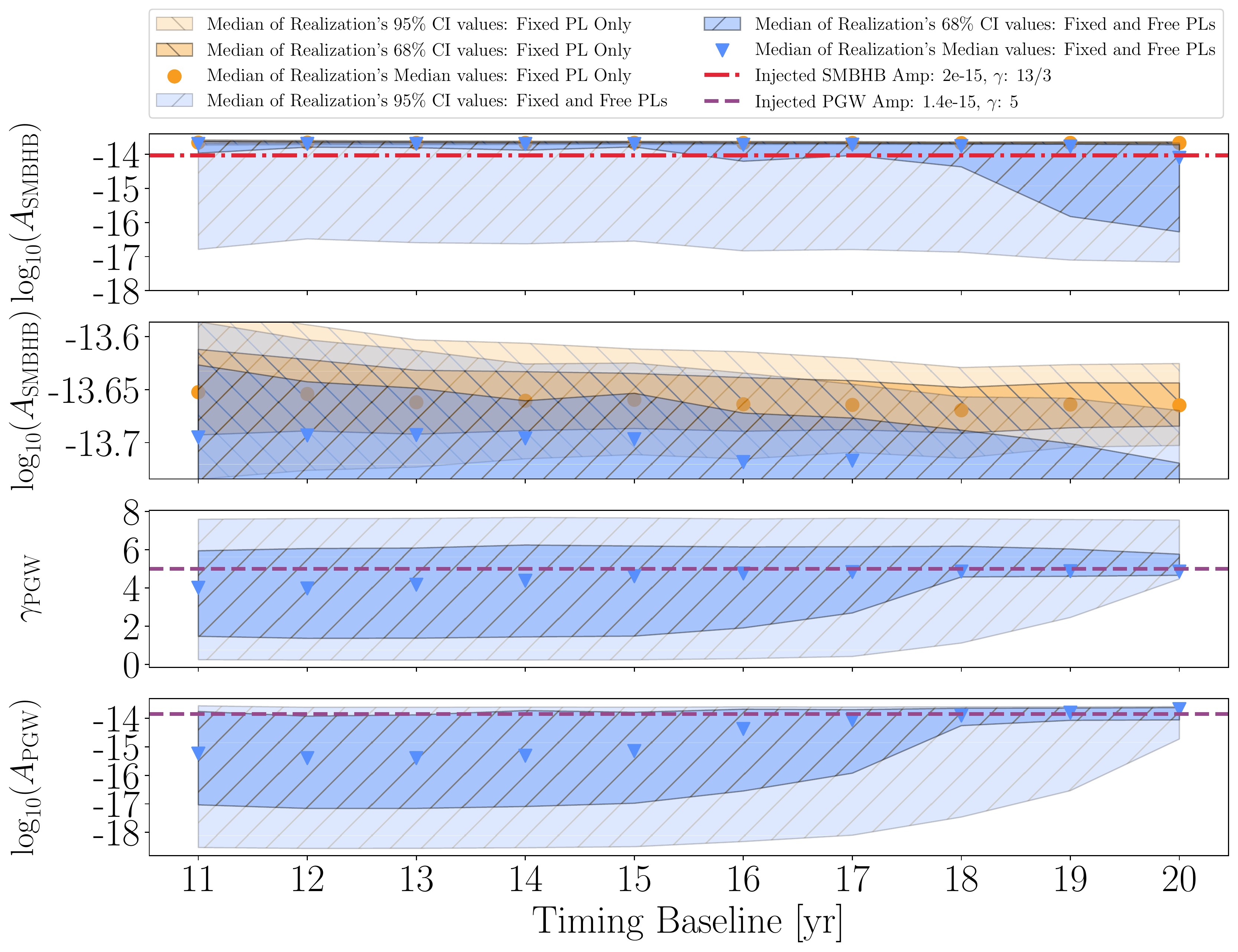}
        \caption{Hypermodel A on an injected density ratio of $\Omega_{\mathrm{PGW}}/\Omega_{\mathrm{SMBHB}} = 0.5$ corresponding to a SMBHB GWB ($\alpha=-2/3$, $\gamma=13/3$) at an amplitude of $\mathrm{A}_{\mathrm{SMBHB,Inj}} = 2\times10^{-15}$ and a PGW GWB ($\alpha=-1$, $\gamma=5$) at $\mathrm{A}_{\mathrm{PGW,Inj}} =1.4\times10^{-15}$ both at $f_\mathrm{ref}=1/\mathrm{yr}$.
        All parameter amplitude panels corresponding to the re-parameterized posteriors in terms of a lower reference frequency of $f_\mathrm{ref}=1/(10\mathrm{yr})$.
        The second from the top plot more closely examines the narrow region of the fixed PL only signal model shown in the top panel.
        }
        \label{fig:Om05modelcomp}
    \end{figure}

    \input{Tables/Hypermodel_A_info_vertical}
    
    In figure \ref{fig:Om05modelcomp}, we show the results of using a hypermodel A from section \S\ref{par:fixedvsfree} on the loudest injection with $\Omega_{\mathrm{PGW}}/\Omega_{\mathrm{SMBHB}} = 0.5$ over the course of our 20 year, 50 realization simulated datasets.
    We represent the 50 realizations at each time slice by taking the medians and confidence intervals (CIs) for individual realization posteriors, then taking the median of those values over all of the realizations.
    As our analyses use the typical reference frequency of $f_\mathrm{ref}=1/\mathrm{yr}$, and for this study we are concerned only with separating the spectral indices at lower frequencies, we then re-parameterize the posteriors in terms of a lower reference frequency of $f_\mathrm{ref}=1/(10\mathrm{yr})$.
    We quote errors at frequencies closer to those of concern to more accurately represent the recovered parameter space.
    We report recovered values at both $f_\mathrm{ref}=1/\mathrm{yr}$ and $f_\mathrm{ref}=1/(10\mathrm{yr})$ in table \ref{tab:Hypermodel_A_info}.
    It is clear that the recovered amplitudes of the SMBHB GWB in the fixed PL only model overestimate the injected value (see the inset in figure \ref{fig:Om05modelcomp}). 
    The majority of the SMBHB GWB posterior space for the fixed PL combined with a free PL prefers a higher amplitude for the injected value until the second process' constraints begin to improve. 
    At the 20 year mark, the median value of all 50 realizations lies around the injected values (see table \ref{tab:Hypermodel_A_info}).
    We suspect that the overestimation of the SMBHB GWB is due in part to exchanging power with the PGW GWB injection as the amplitudes of each are covariant with each other.
    
    \begin{figure}[!htbp]
        \includegraphics[width=\columnwidth]{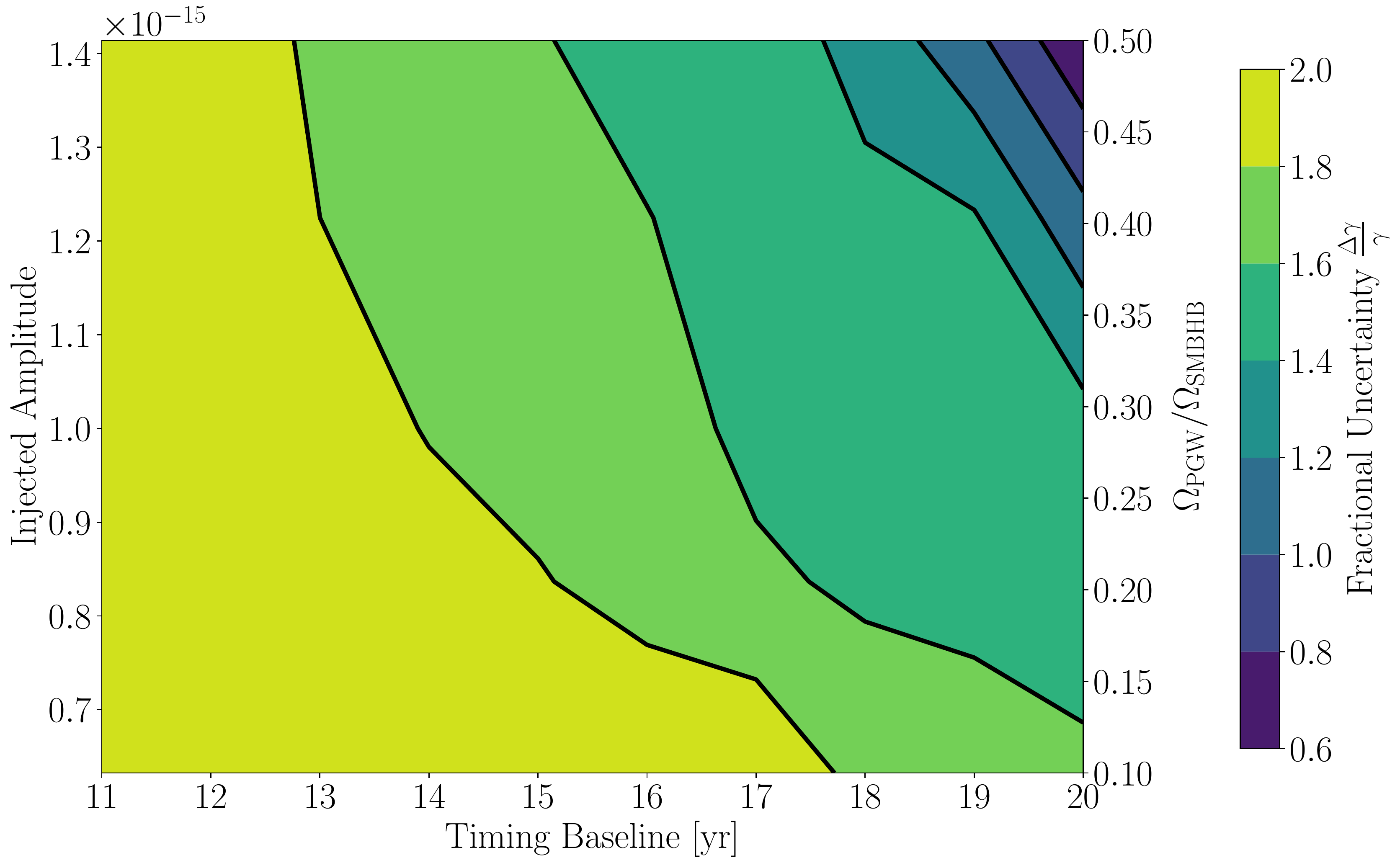}
        \caption{Fractional uncertainties on the spectral index for 50 realizations of an additional process on top of a SMBHB GWB fixed PL with respect to injected strain amplitudes at $f_\mathrm{ref}=1/\mathrm{yr}$ (left axis) and GW density fraction (right axis) versus the timing baseline of the PTA. 
        }
        \label{fig:fracuncertgamma30fixedand30free}
    \end{figure}
    
\begin{figure}[!htbp]
        \includegraphics[width=\columnwidth]{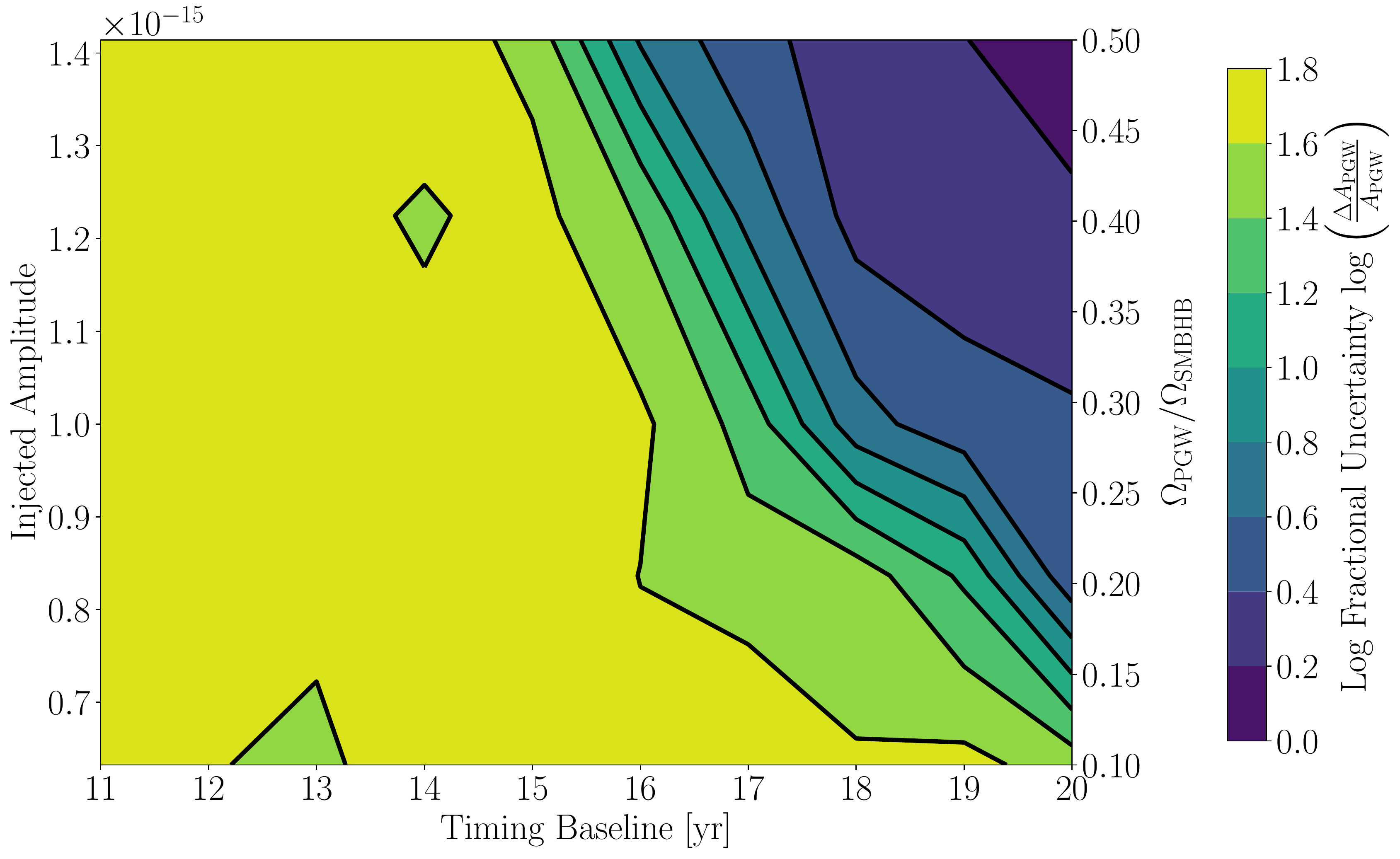}
        \caption{Logarithmic fractional uncertainties on the amplitude at $f_\mathrm{ref}=1/(10\mathrm{yr})$ for 50 realizations of an additional process on top of a SMBHB GWB fixed PL with respect to injected strain amplitudes at $f_\mathrm{ref}=1/\mathrm{yr}$ (left axis) and GW density fraction (right axis) versus the timing baseline of the PTA.
        }
        \label{fig:fracuncertampother30fixedand30free}
    \end{figure}

    To determine the evolution of the constraints on the parameters of another signal in the data, we use the fractional parameter uncertainty $\Delta X/X$, where $X$ is the median measured value and $\Delta X$ is the $95\%$ CI uncertainty of the relevant parameter.
    We show the results of this in figures \ref{fig:fracuncertgamma30fixedand30free} and \ref{fig:fracuncertampother30fixedand30free} for the spectral index and amplitude, respectively.
    To keep the comparisons between figures simple, we report the fractional uncertainties for figure \ref{fig:fracuncertampother30fixedand30free} with reference frequencies for $f_\mathrm{ref}=1/(10\mathrm{yr})$, yet keep the injected amplitude labels for a reference frequency of $f_\mathrm{ref}=1/\mathrm{yr}$.
    
    In both cases, the parameter becomes more constrained as the timing baseline increases and as the strength of the PGW GWB increases.
    The spectral index of the PGW GWB achieves a minimum fractional uncertainty of $64\%$ at the 20-year mark for a GW density ratio compared to the SMBHB GWB of $0.5$.
    Since the measured spectral index will be the primary identifying criterion of an additional GWB, it is necessary to constrain the parameter to a high degree.
    
    The fractional uncertainty of the PGW GWB amplitude is much less constrained that the background's spectral index even after shifting the reference frequency to $f_\mathrm{ref}=1/(10\mathrm{yr})$.
    Over the simulated 20 years in this study, we expect the amplitude of an additional process below a foreground GWB to have a fractional uncertainty of over $110\%$ at $f_\mathrm{ref}=1/(10\mathrm{yr})$ for all injected amplitudes.
    Much like the predictions in \cite{Pol2021} however, we expect the constraints to continue to improve as the PTA accrues more observing time.
    
    \begin{figure*}
        \centering
        \includegraphics[width=\textwidth]{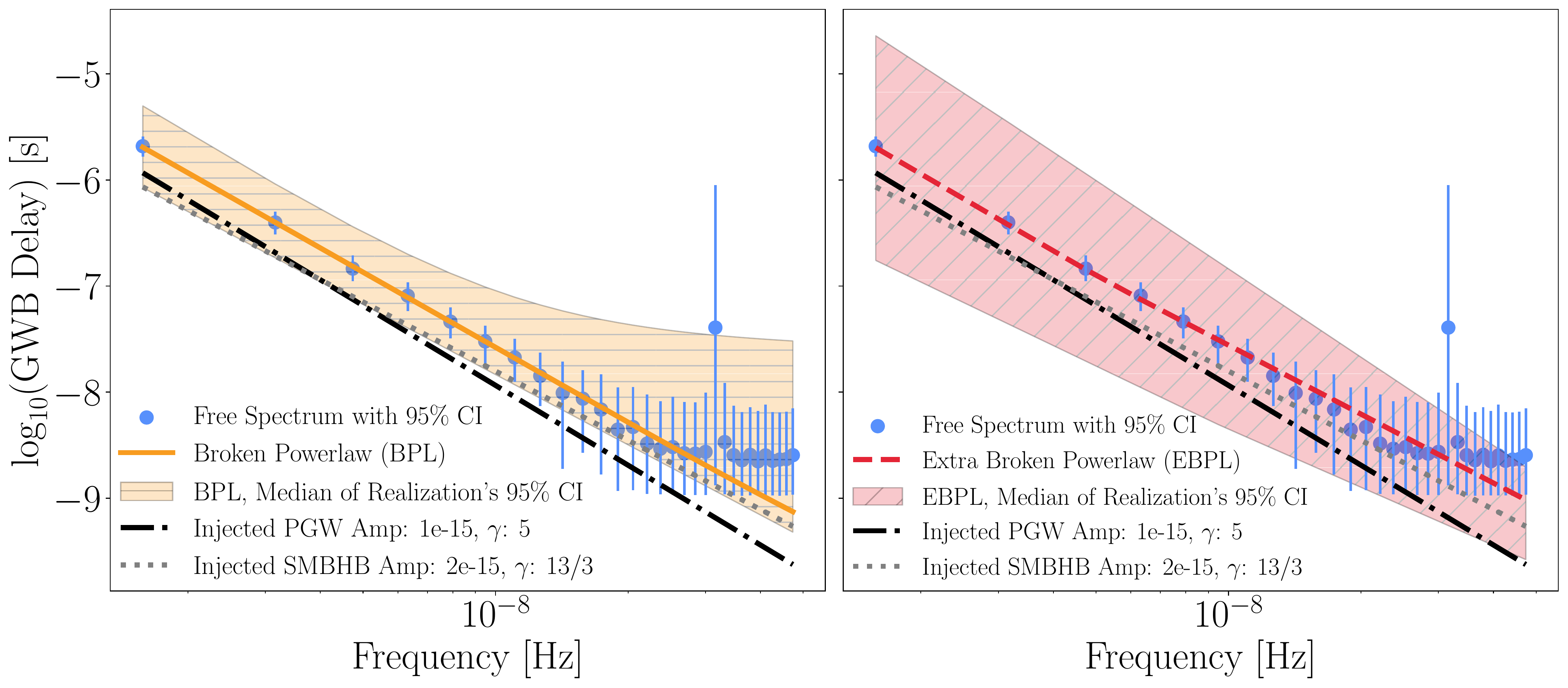}
        \caption{Comparison of medians and $95\%$ CIs in terms of the logarithmic GWB delay (eqn. \ref{eq:GWB_res_delay}) of three models for 50 realizations: the free spectrum (blue dots and lines) in both panels, the BPL in the left panel, with its median (solid orange line) and $95\%$ CI (vertically striped orange shaded region), and the EBPL in the right panel, with its median (dashed red line) and $95\%$ CI (horizontally striped red shaded region).
        We compare each model to the injected SMBHB GWB ($\alpha=-2/3$, $\gamma=13/3$) at our nominal amplitude of $2\times10^{-15}$ at $f_\mathrm{ref}=1/\mathrm{yr}$ in the dotted grey line and the injected PGW GWB ($\alpha=-1$, $\gamma=5$) of density ratio $\Omega_{\mathrm{PGW}}/\Omega_{\mathrm{SMBHB}} = 0.25$ in the dashed-dotted black line. 
        }
        \label{fig:dgwb_EBPL_BPL_PL_modelcomp}
    \end{figure*}
    
    In figure \ref{fig:dgwb_EBPL_BPL_PL_modelcomp}, we show the results of the EBPL, the BPL, and the free spectrum analyses performed on the $\Omega_{\mathrm{PGW}}/\Omega_{\mathrm{SMBHB}} = 0.25$ injection.
    We vary all parameters present in equations \ref{eq:brokenplaw} and \ref{eq:extrabrokenplaw} for the BPL and EBPL, respectively.
    Both the EBPL and the BPL presented here were analyzed with the hypermodel D and hypermodel E framework, respectively.
    In all the comparisons, we find that the simplest model is preferred, e.g. the 30 frequency free PL is preferred over the BPL and EBPL.
    It is clear from the medians in figure \ref{fig:dgwb_EBPL_BPL_PL_modelcomp} (the solid orange and dashed red lines) that a single PL without breaks in frequency nor changes in spectral index is the dominant feature.

    The medians also trace the free spectral model that represents the power at each frequency.
    Both the free spectrum and medians of the PLs seem to over-estimate the amplitude of the injected GWBs.
    This was already evident in our earlier analysis shown in figure \ref{fig:Om05modelcomp} where the SMBHB GWB fixed PL recovers a higher amplitude than injected.
    This could pose a problem when attempting to recover the amplitude of multiple backgrounds (see section \S\ref{sec:Discussion} for further discussion).
    
    In an attempt to rule out convergence issues causing the lack of preference for break frequencies where the steeper PGW GWB spectrum dominates and at higher frequencies where the WN dominates, we restricted the ranges at which the break frequencies occur.
    The low frequency break-point is restricted to below $10^{-8}~\mathrm{Hz}$, while the high frequency break occurs above there.
    Despite these restrictions, there is no preferred break frequency in either the high or low frequency regimes.

    While there is still evidence that there is a break frequency in some realizations based on the 95\% CI of the BPL in figures \ref{fig:dgwb_EBPL_BPL_PL_modelcomp} and \ref{fig:sgwb_EBPL_BPL_PL_modelcomp}, the medians of all 50 realizations do not prefer a break in the PL. 
    Even for the case of the EBPL, the 95\% CI is still consistent with the 95\% CI of the free spectral model except for the highest few frequencies in the single PGW GWB injection shown in figure \ref{fig:sgwb_EBPL_BPL_PL_modelcomp}. 
    Since the free spectral model only encompasses the RN at each individual frequency, it has no assumption of the spectrum’s shape.
    The BPL and EBPL require more parameters than a simple free PL, but there seems to be little evidence over the whole spectrum to require the additional parameters.

    We suspect that the injected GWBs together are strong enough to prefer inclusion of all 30 frequencies in a simple free PL despite the WN at higher frequencies.
    Similarly, no low-frequency break is preferred for the EBPL.
    We suspect this is because of the closeness in spectral indices between $\gamma=5$ and $\gamma=13/3$, but expect with baselines of greater than 20 years, evidence for the low-frequency break will increase.
    
    \begin{figure}[!htbp]
        \includegraphics[width=\columnwidth]{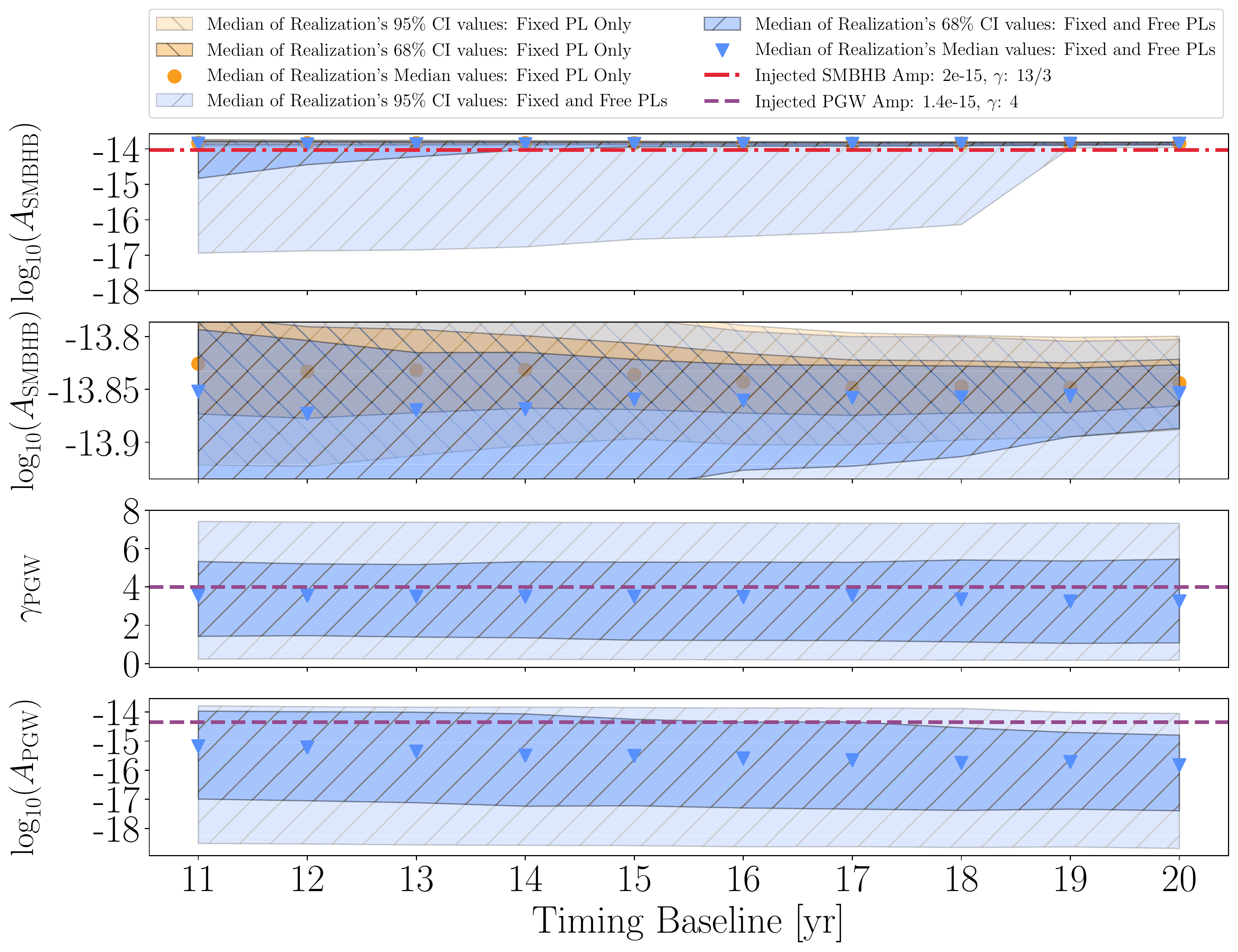}
        \caption{Hypermodel A on an injected density ratio of $\Omega_{\mathrm{PGW}}/\Omega_{\mathrm{SMBHB}} = 0.5$ corresponding to a SMBHB GWB ($\alpha=-2/3$, $\gamma=13/3$) at an amplitude of $\mathrm{A}_{\mathrm{SMBHB,Inj}} = 2\times10^{-15}$ and a PGW GWB ($\alpha=-1/2$, $\gamma=4$) at $\mathrm{A}_{\mathrm{PGW,Inj}} =1.4\times10^{-15}$ both at $f_\mathrm{ref}=1/\mathrm{yr}$.
        All parameter amplitude panels corresponding to the re-parameterized posteriors in terms of a lower reference frequency of $f_\mathrm{ref}=1/(10\mathrm{yr})$.
        The second from the top plot more closely examines the narrow region of the fixed PL only signal model shown in the top panel.
        }
        \label{fig:Om05modelcomp_gam4}
    \end{figure}
    
    Since there is the possibility of GWBs containing shallower spectral indices than the SMBHB GWB (e.g. a population of eccentric binaries \citep{Taylor2017,Chen2019}, or PGWs), we inject a $\Omega_{\mathrm{PGW}}/\Omega_{\mathrm{SMBHB}} = 0.5$ PGW GWB with a spectral index of $\alpha=-1/2$ ($\gamma=4$).
    The resulting medians and CIs of a hypermodel A for 50 realizations is shown in figure \ref{fig:Om05modelcomp_gam4}.
    We find that even using the largest amplitude injection from our analysis of the steeper PGW GWB, the hypermodel weakly prefers only an SMBHB GWB. 
    While the model with an additional free PL has a median around the injected spectral index value, there is not much improvement in the constraints on the injected PGW GWB parameters.
    
\subsection{Single GWB Sources}
    \label{subsec:SingleGWBSources}
    In order to verify the veracity of future potential claims of detection of two backgrounds, we investigate the effect of using the same analysis methods on a single background injection.
    \cite{Pol2021} investigates the detectability of a single SMBHB GWB with different spectral components at lower frequencies based on different population models, and find that they can differentiate different spectra at around the 17-year slice.
    Thus we inject only a background from PGWs at an amplitude of $A_{\mathrm{yr}}=1\times10^{-15}$ and confirm our methods of detecting two backgrounds are sufficiently rigorous to prefer a PGW GWB and rule out the SMBHB GWB assumed to be in the data. 
    
    \begin{figure}[!htbp]
        \includegraphics[width=\columnwidth]{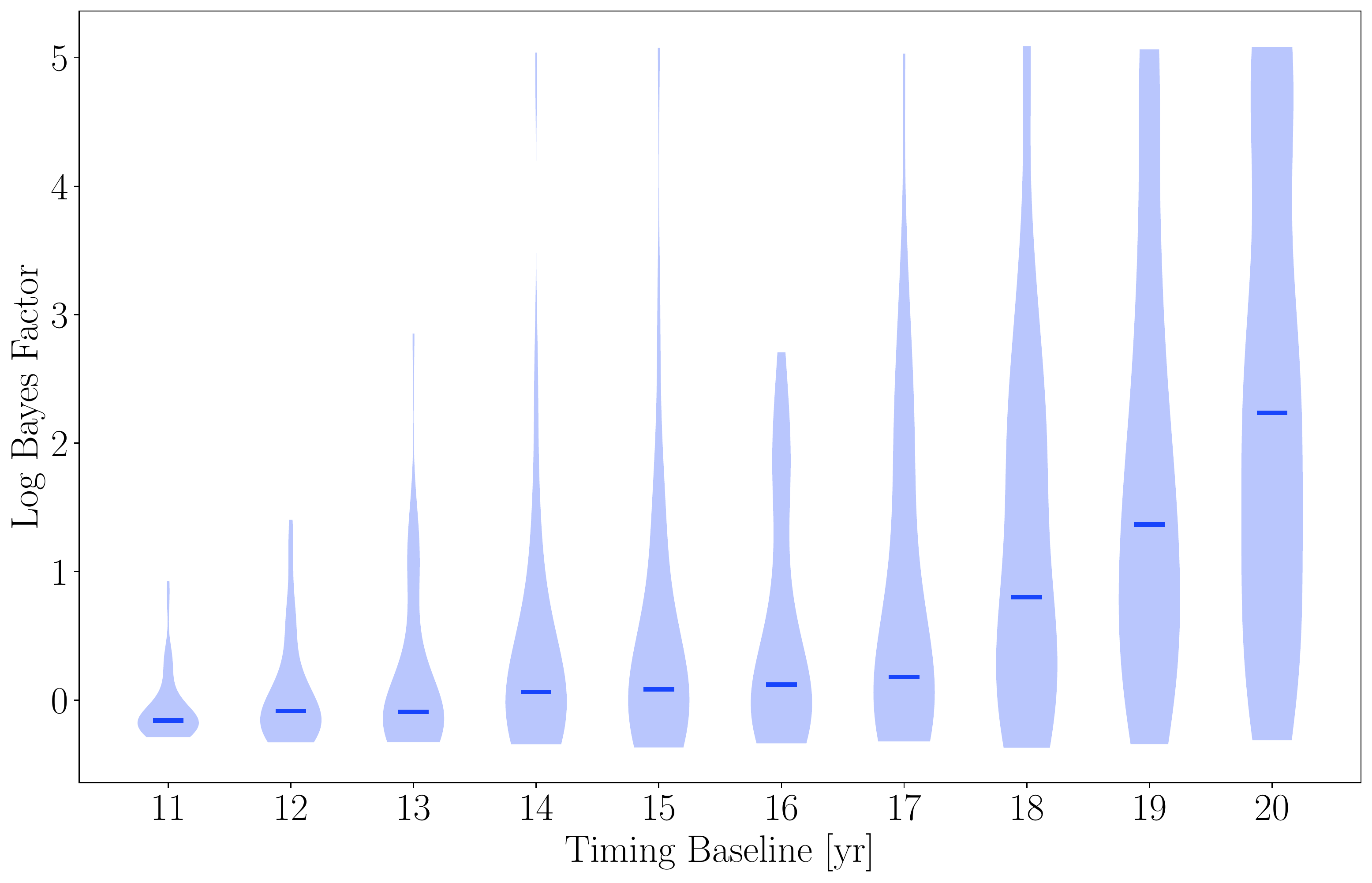}
        \caption{Medians (blue lines) and distributions of $\mathrm{log}_{10}$ Bayes' factors for 50 realizations of a hypermodel A analysis versus the timing baseline of the PTA. 
        Only a PGW GWB injected in this dataset at an amplitude of $A_{\mathrm{PGW}}=1\times10^{-15}$ and spectral index $\alpha=-1$ ($\gamma=5$) and thus these BFs represent the odds of a separate free PL on top of a fixed PL, or just the fixed PL to show the robustness of our previous analyses in the presence of only a PGW GWB.
        }
        \label{fig:sgwb_BFs}
    \end{figure}
    
    In figure \ref{fig:sgwb_BFs} we show the BFs for the hypermodel A where we examine the evidence of including a free PL at indices other than at $\gamma=13/3$. 
    It is clear that an additional process begins to be preferred in all baselines after around 18 years with the strength of preference growing with the timing baseline.
    Thus if we use a hypermodel A on data that contains only one signal at a different spectral index than the fixed PL, the inclusion of the free PL can indicate that the signal has been mis-modeled.
    This is a positive indicator that the hypermodel framework we use can differentiate between a signal at the fixed PL's spectral index and one with a similar spectral index.
    
    \begin{figure}[!htbp]
        \includegraphics[width=\columnwidth]{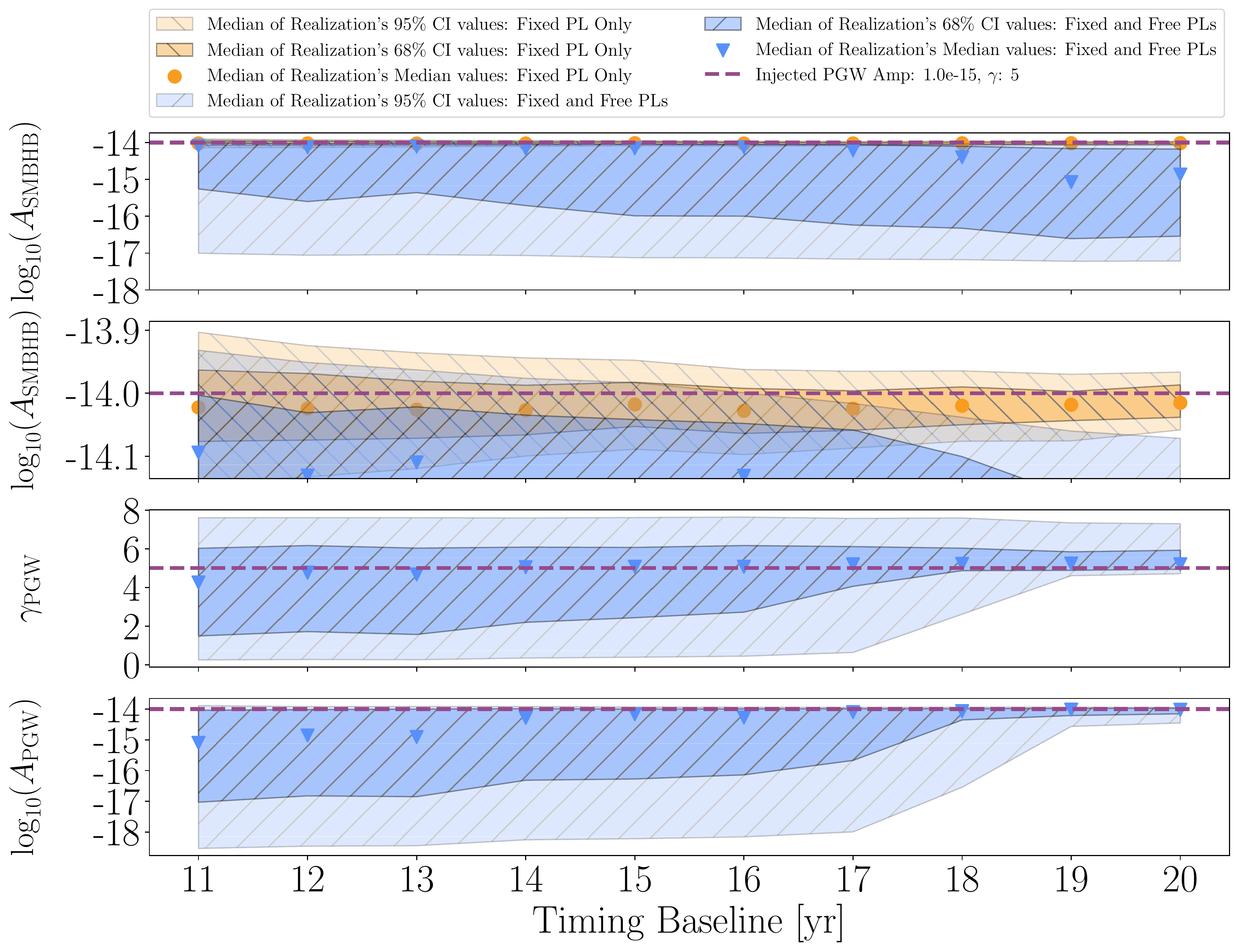}
        \caption{Hypermodel A on a single injection in this dataset with an amplitude of $\mathrm{A}_{\mathrm{PGW,Inj}} =1\times10^{-15}$ at $f_\mathrm{ref}=1/\mathrm{yr}$ with a spectral index of $\alpha=-1$ ($\gamma=5$).
        All parameter amplitude panels corresponding to the re-parameterized posteriors in terms of a lower reference frequency of $f_\mathrm{ref}=1/(10\mathrm{yr})$.
        The second from the top plot more closely examines the narrow region of the fixed PL only signal model shown in the top panel.
        }
        \label{fig:sgwb_Om025modelcomp}
    \end{figure}
    
    Despite including a signal model not present in the data, we are able to constrain the single injected GWB to a fractional uncertainty of $50\%$ in the spectral index and $84\%$ in amplitude at $f_\mathrm{ref}=1/(10\mathrm{yr})$, shown in figure \ref{fig:sgwb_Om025modelcomp}.
    By including the incorrect model however, we find again that some of the power of the injected PGW GWB is absorbed into the SMBHB fixed spectral model.
    The effect decreases as the baseline time extends due to the push into lower frequencies and thus greater resolution to differentiate between the SMBHB and the steeper PGW GWB spectral index.
    
    \begin{figure*}
        \centering
        \includegraphics[width=\textwidth]{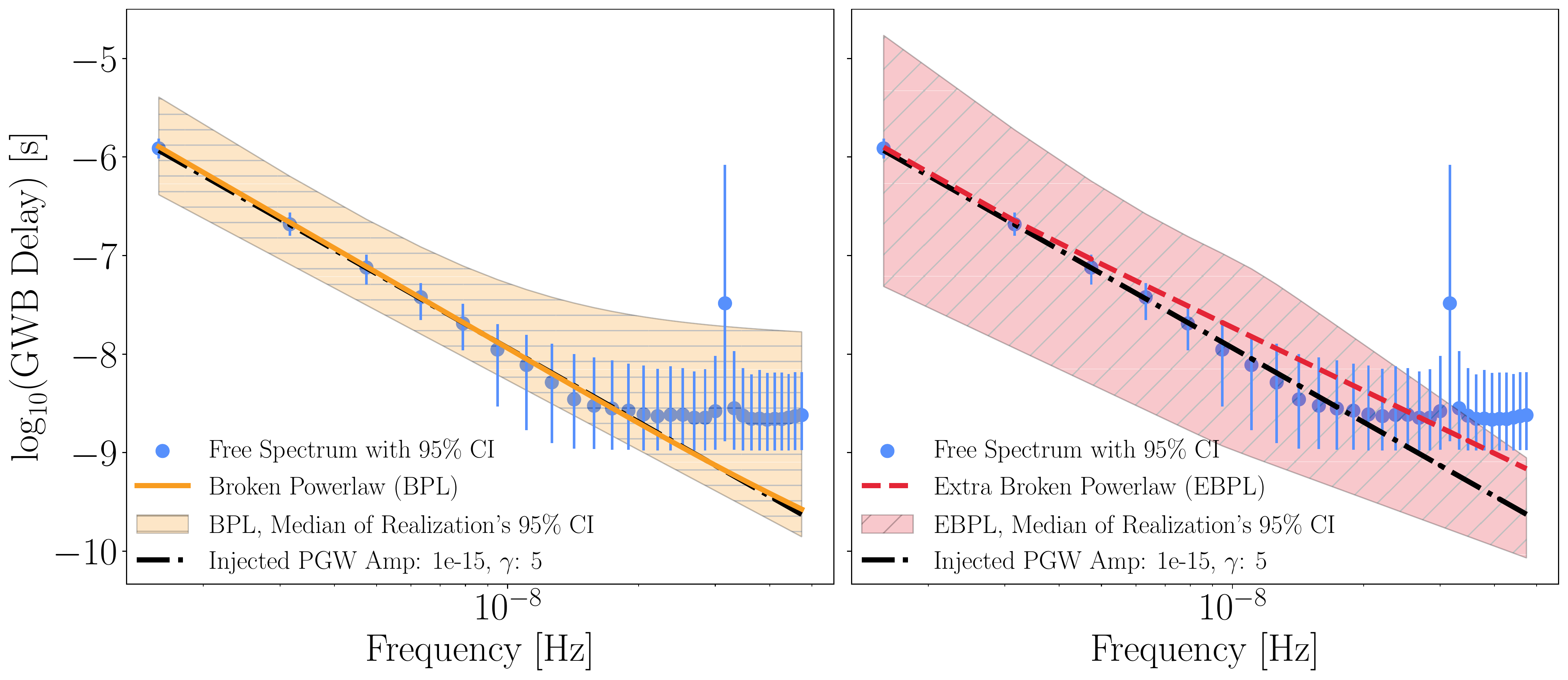}
        \caption{Comparison of medians and $95\%$ CIs of the logarithmic GWB delay of three models for 50 realizations as they evolve over time: the free spectrum (blue dots and lines), the BPL median (solid orange line) and $95\%$ CI (vertically striped orange shaded region), and the EBPL median (dashed red line) and $95\%$ CI (horizontally striped red shaded region).
        We compare each model to the injected PGW GWB ($\alpha=-1$, $\gamma=5$) of amplitude $A_{\mathrm{PGW}}=1\times10^{-15}$ at $f_\mathrm{ref}=1/\mathrm{yr}$ in the dashed-dotted black line.
        }
        \label{fig:sgwb_EBPL_BPL_PL_modelcomp}
    \end{figure*}
    
    We now perform the same analyses as in figure \ref{fig:dgwb_EBPL_BPL_PL_modelcomp} on the single injection of a PGW GWB with an amplitude of $10^{-15}$ at $f_\mathrm{ref}=1/\mathrm{yr}$ in figure \ref{fig:sgwb_EBPL_BPL_PL_modelcomp}. 
    We use the same hypermodel scheme to mimic our analysis with the double injection where we compare the more complicated broken PLs to a simpler free PL.
    Again we find that the simplest model is preferred: a single free PL that includes all 30 frequencies used.
    In this single injection analysis, we find that the free spectrum and the PLs well characterize the injected spectrum and do not over-estimate the amplitudes.
    This lends support to the need for multiple models when searching for more than one background, but highlights the difficulty of correctly parameterizing each GWB.

%% file: Tables/Hypermodel_A_info_vertical.tex
\begin{table*}[!htbp]
    \centering
    \caption{Hypermodel A Results}
    \begin{tabular}{@{} c|cccccccc @{}}
        \toprule
            $\frac{\Omega_{\mathrm{PGW},\alpha=\text{--}1}}{\Omega_{\mathrm{SMBHB}}}$ & & \multirow{2}{*}{0.5} & \multirow{2}{*}{0.375} & \multirow{2}{*}{0.25} & \multirow{2}{*}{0.175} & \multirow{2}{*}{0.1} & $\frac{\Omega_{\mathrm{PGW},\alpha=\text{--}1/2}}{\Omega_{\mathrm{SMBHB}}}$ & \multirow{2}{*}{0.5} \\
            $f_\mathrm{ref}=\frac{1}{\mathrm{yr}}$ & & & & & & & $f_\mathrm{ref}=\frac{1}{\mathrm{yr}}$ & \\
        \midrule
            \multirow{2}{*}{$\mathrm{log}_{10}(\mathrm{A}_{\mathrm{PGW,Inj}})$} & $f_\mathrm{ref}=\frac{1}{\mathrm{yr}}$ & $\text{--}14.85$ & $\text{--}14.91$ & $\text{--}15.0$ & $\text{--}15.08$ & $\text{--}15.20$ & & $\text{--}14.85$ \\
            
            & $f_\mathrm{ref}=\frac{1}{10\mathrm{yr}}$ & $\text{--}13.85$ & $\text{--}13.91$ & $\text{--}14.0$ & $\text{--}14.10$ & $\text{--}14.20$ & & $\text{--}14.35$ \\
        \midrule
        \midrule
            Signal Model 1 & & & & & & & \\
        \midrule
        \midrule
            \multirow{2}{*}{$\mathrm{log}_{10}(\mathrm{A}_{\mathrm{fixed}})$} & $f_\mathrm{ref}=\frac{1}{\mathrm{yr}}$ & $\text{--}14.8^{+0.4}_{\text{--}3.1}$ & $\text{--}14.7^{+0.3}_{\text{--}3.0}$ & $\text{--}14.5^{+.1}_{\text{--}3.0}$ & $\text{--}14.5^{+0.1}_{\text{--}2.8}$ & $\text{--}14.54^{+0.08}_{\text{--}2.07}$ & & $\text{--}14.54^{+0.09}_{\text{--}0.07}$ \\
            
            & $f_\mathrm{ref}=\frac{1}{10\mathrm{yr}}$ & $\text{--}13.66_{\text{--}0.04}^{+0.04}$ & $\text{--}13.7_{\text{--}0.04}^{+0.04}$ & $\text{--}13.74_{\text{--}0.04}^{+0.04}$ & $\text{--}13.77_{\text{--}0.04}^{+0.04}$ & $\text{--}13.82_{\text{--}0.04}^{+0.04}$ & & $\text{--}13.86_{\text{--}0.09}^{+0.07}$ \\
        \midrule
        \midrule
        Signal Model 2 & & & & & & & \\
        \midrule
        \midrule
            \multirow{2}{*}{$\mathrm{log}_{10}(\mathrm{A}_{\mathrm{fixed}})$} & $f_\mathrm{ref}=\frac{1}{\mathrm{yr}}$ & $\text{--}14.8^{+0.4}_{\text{--}3.1}$ & $\text{--}14.7^{+0.3}_{\text{--}3.0}$ & $\text{--}14.5^{+.1}_{\text{--}3.0}$ & $\text{--}14.5^{+0.1}_{\text{--}2.8}$ & $\text{--}14.54^{+0.08}_{\text{--}2.07}$ & & $\text{--}14.54^{+0.09}_{\text{--}0.07}$ \\
            
            & $f_\mathrm{ref}=\frac{1}{10\mathrm{yr}}$ & $\text{--}14.1_{\text{--}3.06}^{+0.43}$ & $\text{--}13.8_{\text{--}3.2}^{+0.1}$ & $\text{--}13.79_{\text{--}3.11}^{+0.07}$ & $\text{--}13.83_{\text{--}2.83}^{+0.08}$ & $\text{--}13.85_{\text{--}2.09}^{+0.05}$ & & $\text{--}13.87_{\text{--}0.09}^{+0.07}$ \\
        \midrule
            \multirow{2}{*}{$\mathrm{log}_{10}(\mathrm{A}_{\mathrm{free}})$} & $f_\mathrm{ref}=\frac{1}{\mathrm{yr}}$ & $\text{--}14.7^{+0.2}_{\text{--}2.0}$ & $\text{--}14.7^{+0.3}_{\text{--}2.2}$ & $\text{--}15.5^{+1.0}_{\text{--}2.2}$ & $\text{--}15.4^{+0.9}_{\text{--}2.4}$ & $\text{--}15.8^{+1.3}_{\text{--}2.0}$ & & $\text{--}15.7^{+1.4}_{\text{--}2.2}$ \\
            
            & $f_\mathrm{ref}=\frac{1}{10\mathrm{yr}}$ & $\text{--}13.68_{\text{--}1.05}^{+0.09}$ & $\text{--}13.9_{\text{--}2.6}^{+0.2}$ & $\text{--}14.1_{\text{--}3.8}^{+0.4}$ & $\text{--}14.2_{\text{--}4.0}^{+0.5}$ & $\text{--}15.3_{\text{--}3.2}^{+1.5}$ & & $\text{--}15.4_{\text{--}3.2}^{+1.4}$ \\
        \midrule
            \multirow{2}{*}{$\frac{\Delta \mathrm{log}_{10}(\mathrm{A}_{\mathrm{free}})}{\mathrm{log}_{10}(\mathrm{A}_{\mathrm{free}})}$} & $f_\mathrm{ref}=\frac{1}{\mathrm{yr}}$ & $1.6$ & $3.7$ & $11$ & $15$ & $34$ & & $79$\\
            
            & $f_\mathrm{ref}=\frac{1}{10\mathrm{yr}}$ & $1.1$ & $1.8$ & $2.7$ & $2.8$ & $32$ & & $60$\\
        \midrule
            $\gamma_{\mathrm{free}}$ & & $5.0^{+2.6}_{\text{--}0.5}$ & $5^{+3}_{\text{--}2}$ & $5^{+2}_{\text{--}5}$ & $5^{+3}_{\text{--}5}$ & $5^{+3}_{\text{--}5}$ & & $3^{+4}_{\text{--}3}$ \\
        \midrule
            $\frac{\Delta \gamma_{\mathrm{free}}}{\gamma_{\mathrm{free}}}$ & & $0.64$ & $1.1$ & $1.5$ & $1.8$ & $1.6$ & & $2.2$ \\
        \midrule
        \midrule
            $\mathrm{BF}$ & & $16$ & $7$ & $1.7$ & $1.5$ & $0.7$ & & $0.5$ \\
        \midrule
            $\mathrm{High~BFs}$ & & $22\%$ & $12\%$ & $10\%$ & $8\%$ & $6\%$ & & $0\%$ \\
        \bottomrule

    \end{tabular}
    \tablecomments{Hypermodel A results on six PGW GWB injected amplitudes all with an injected SMBHB GWB of $A_{\mathrm{SMBHB}}=2\times10^{\text{--}15}$ ($\mathrm{log}_{10}(A_{\mathrm{SMBHB}})=\text{--}14.70$) at $f_\mathrm{ref}=\frac{1}{\mathrm{yr}}$, or $A_{\mathrm{SMBHB}}=9.28\times10^{\text{--}15}$  ($\mathrm{log}_{10}(A_{\mathrm{SMBHB}})=\text{--}14.03$) at $f_\mathrm{ref}=\frac{1}{10\mathrm{yr}}$ with a spectral index of $\gamma=13/3$ for a 20 year PTA timing baseline.
    The $\mathrm{BF}$ represents the median of 50 realizations' median Bayes' factors.
    The final row of $\mathrm{High~BFs}$ represents the percentage of realizations with $\mathrm{log}_{10}(\mathrm{BF})>3$.
    The fractional parameter uncertainty is $\Delta X/X$, where $X$ is the median measured value and $\Delta X$ is the 95\% CI uncertainty of the relevant parameter.
    All other recovered numbers represent the medians of 50 realizations' medians and 95\% CI.}
    \label{tab:Hypermodel_A_info}
\end{table*}

%% file: Section_Discussion.tex
\subsection{Detecting Multiple Backgrounds}
    \paragraph{Separability timescale}
        Over the course of a 20-yr  timing baseline, we expect the evidence for a single GWB to grow rapidly.
        The origin of a GWB will continue to become more evident, from a population of SMBHBs, PGWs, cosmic strings, other exotic sources, and their combinations.
        PTAs are already on track to reach sensitivities that are more than sufficient to detect the SMBHB background within several years \citep{Pol2021}. 
        
        For the case of multiple backgrounds, the median BF of a marginally steeper process of $\alpha=-1$ ($\gamma=5$) remains low over 50 realizations, however in $10\%$ of realizations the evidence for an additional process is highly significant by year 20 even for a moderately strong signal of $\Omega_{\mathrm{PGW},\alpha=-1}/\Omega_{\mathrm{SMBHB}} = 0.25$. 
        For a shallower process of $\alpha=-1/2$ ($\gamma=4$), however, we expect more time to be needed to separate out a moderately strong background from a foreground signal.
        
    \paragraph{How well can we distinguish between two backgrounds?}
        Based on our simulations, even if there is strong evidence for a second GWB (BFs $>1000$), accurate parameter recovery is more difficult.
        Even after 20 years  with a moderately strong signal of $\Omega_{\mathrm{PGW},\alpha=-1}/\Omega_{\mathrm{SMBHB}} = 0.5$ using the methods used here we can only constrain the PL parameters to a fractional uncertainty of $64\%$ on $\gamma_{\mathrm{PGW}}$ and $110\%$ on $\mathrm{log}_{10}(A_{\mathrm{PGW}})$ for a PGW GWB with spectral index of $\alpha=-1$ ($\gamma=5$). 
        There are compounding problems with over-estimating the SMBHB background as well.
        The good news is that the evidence for an additional process (through a hypermodel A analysis) grows at an increasing rate and the constraints seem to decrease by a few to tens of percents per year. 
        For a shallower spectrum, we cannot put constraints on either parameter until the baseline is at least longer than 20 yr.
        Applying frequentist methods like those developed in \cite{Parida2016,Parida2019} for ground-based and \cite{Boileau2021} for space-based interferometric GW observatories to separate multiple GW signals into their components with distinct spectral indices, to PTAs will continue to improve prospects of detection and separation of multiple GWBs.

    \paragraph{Protocol for detecting multiple backgrounds}
        Because PTAs will have continually increasing sensitivity to GWBs from astrophysical and cosmological origin, it is important to have strategies for searching for multiple signals.
        Using the exploratory study presented here, we outline the first protocol for initial searches for multiple GWBs in PTA datasets using Bayesian methods.
        The methods listed can and should be adapted to individual datasets and searches, but provide a guide to initial forays.
        \begin{enumerate}
            \item Use a free spectral model to analyze the power at each frequency without a PL assumption.
            \item Find out if there is one or two distinct break frequencies with a BPL and an EBPL.
            This should be done in the hypermodel framework to not force a break when there is no (Bayesian) evidence for one.
            \item Use a hypermodel A to search for excess power.
            This could be done by searching only the frequencies below the break(s) in the BPL and/or EBPL analysis.
            Searching only the lowest frequencies is particularly sensitive to steeper spectrums with relatively low amplitudes at $f_\mathrm{ref}=1/\mathrm{yr}$ since it will not be significantly affected by higher frequency WN.
            Searching over 30 (or more) frequencies is also a good idea, especially if the broken power-laws did not prefer break frequencies.
            \item If there is relatively strong evidence for another stochastic process:
            \begin{enumerate}
                \item Perform a hypermodel C analysis to confirm if there is evidence for the fixed PL at the spectral index searched for in the hypermodel A analysis.
                \item Perform a hypermodel B search at the recovered spectral index from the hypermodel A free PL to help to reduce the parameter space.
            \end{enumerate}
            \item Then one can move on to different combinations of hypermodel A using the recovered index as the fixed process to confirm that multiple processes at the predicted spectral indices truly have (Bayesian) support.
        \end{enumerate}
        
\subsection{Mischaracterizing a GWB}
    \paragraph{A single two-signal model with two injected backgrounds}
        In the hypermodel A setup, we find that the amplitude of the stronger signal is biased higher compared to the second, lower signal.
        We show that with more time, the covariance between the two can be mitigated and thus the louder signal's amplitude will decrease as the other increases.
        This assumes a steeper index, lower amplitude signal as the low frequency end will eventually dominate at the lowest frequencies.
        For a shallower process, the more time observed, the better separability we can achieve, but whether two models would be preferred over one with a higher amplitude is work left for later. 
            
    \paragraph{A single-signal model with two injected backgrounds}
        Up to the 20-yr baseline, we find that using a single EBPL, BPL, or PL all seem to prefer a single free PL even when injecting two GWBs. 
        In all cases, we found the amplitude of the single model recovered a higher value than the injected amplitudes.
        Furthermore, the single free PL model recovered a spectral index centered between the two injected indices of $\gamma=13/3$ and $\gamma=5$.
        The spectral index skew 
        only grew worse as more frequencies were added due to the high frequencies preferring the shallower spectrum, while the lower frequencies were dominated by the $\gamma=5$ process.
        An assumption of a single source in a dataset with multiple GWBs could affect the recovered parameters both in amplitude and spectral index.
        It is strongly recommended to search for excess power due to other sources as even a separate background with $\Omega_{\mathrm{PGW}}/\Omega_{\mathrm{SMBHB}} = 0.1$ can lead to misconstruing the GWB source.
        
    \paragraph{A single two-signal model with only one injected background}
        In the case of using a single two-signal model with only one signal injected, we find that the hypermodel methods we use rule out two signals after 20 years of data and in $14\%$ of realizations the evidence for excluding the extra, fixed process is highly significant by year 18.
        It is clear from the work here that using a two-signal model when only one source is present can be corrected quickly when using Bayesian odds to rule out the incorrect model.
        The erroneous recovered signal appears unconstrained and consistent with zero, which should immediately fall under skepticism if the free PL model has constrained posteriors and are not consistent with zero. 
        
        A hypermodel comparing a single free PL with a fixed PL and free PL could determine whether there is truly evidence for the fixed process.
        These results again highlight the need for multiple checks and axes of analyzing datasets and not just searching for what one predicts to find.
        
\subsection{Conclusions}
    In summary, this work explores the current and future potential for PTAs to detect multiple GWBs.
    We find that within the next two-to-five years PTAs will be capable of determining whether there is a combination of GWBs within their data.
    Should there be evidence for multiple GWBs, constraints will continue to improve on the fractional uncertainty by several to tens of percent each year.
    We also highlight the immediate need to search for multiple GWBs as neglecting to do so when two GWBs are present can artificially inflate the amplitude and skew the spectral index for an assumed single background. 

%% file: biblio.tex
%\newcommand{\aap}{\textit{Astron. Astrophys.}}               %{Astronomy and Astrophysics}
\newcommand{\aass}{\textit{Astron. Astrophys. Supp. Ser.}}       %{Astronomy and Astrophysics Supplement Series}
\newcommand{\aipcp}{\textit{AIP Conf. Proc.}}                     %{AIP Conference Proceedings}
\newcommand{\ajp}{\textit{American Journal of Physics}}         %{American Journal of Physics}
\newcommand{\am}{\textit{Ann. Math.}}                          %{Annals of Mathematics}
\newcommand{\anyas}{\textit{Ann. N. Y. Acad. Sciences}}           %{Annals of the New York Academy of Sciences}
\newcommand{\arnps}{\textit{Annu. Rev. Nucl. Part. Sci.}}         %{Annual Review of Nuclear and Particle Science}
\newcommand{\cqg}{\textit{Class. Quantum Grav.}}                %{Classical and Quantum Gravity}
\newcommand{\cmp}{\textit{Commun. Math. Phys.}}                 %{Communications in Mathematical Physics}
\newcommand{\cpc}{\textit{Computer Physics Communications}}     %
\newcommand{\CUP}{\textit{Cambridge University Press}}          %
\newcommand{\grg}{\textit{Gen. Rel. Grav.}}                     %{General Relativity and Gravitation}
\newcommand{\ijmpa}{\textit{Int. J. Mod. Phys. A}}                %{International Journal of Modern Physics A}
\newcommand{\ijmpd}{\textit{Int. J. Mod. Phys. D}}                %{International Journal of Modern Physics D}
\newcommand{\jap}{\textit{J. App. Phys.}}                       %{Journal of Applied Physics}
\newcommand{\jpcs}{\textit{J. Phys. Conf. Ser.}}                 %{Journal of Physics: Conference Series}
\newcommand{\jmp}{\textit{J. Math. Phys.}}                      %{Journal of Mathematical Physics}
\newcommand{\lnp}{\textit{Lecture Notes in Physics}}            %
\newcommand{\lrr}{\textit{Living Rev. Relativity}}              %{Living Reviews in Relativity}
\newcommand{\mpla}{\textit{Mod. Phys. Lett. A}}                  %{Modern Physics Letters A}
\newcommand{\njp}{\textit{New Jour. Phys.}}                     %{New Journal of Physics}
\newcommand{\OUP}{\textit{Oxford University Press}}             %
\newcommand{\pla}{\textit{Phys. Lett. A}}                       %{Physics Letters A}
\newcommand{\plb}{\textit{Phys. Lett. B}}                       %{Physics Letters B}
\newcommand{\prpt}{\textit{Phys. Rept.}}                         %
\newcommand{\pr}{\textit{Phys. Rev.}}                          %{Physical Review}
\newcommand{\prsla}{\textit{Proc. R. Soc. London Ser. A}}         %{Proceedings of the Royal Society of London Series A}
\newcommand{\ptp}{\textit{Prog. Theor. Phys.}}                  %{Progress of Theoretical Physics}
\newcommand{\ptps}{\textit{Prog. Theor. Phys. Suppl.}}           %{Progress of Theoretical Physics Supplement}
\newcommand{\ptrsl}{\textit{Phil. Trans. R. Soc. Lond.}}          %{Philosophical Transactions of the Royal Society of London}
\newcommand{\ptrsla}{\textit{Phil. Trans. R. Soc. London Ser. A}}  %{Philosophical Transactions of the Royal Society of London Series A}
\newcommand{\pw}{\textit{Physics World}}                       %
\newcommand{\rpp}{\textit{Rep. Prog. Phys.}}                    %{Reports on Progress in Physics}
\newcommand{\WS}{\textit{World Scientific}}                    %

\providecommand{\newblock}{}

%% file: main.bbl
\begin{thebibliography}{30}
\expandafter\ifx\csname natexlab\endcsname\relax\def\natexlab#1{#1}\fi
\providecommand{\eprint}[2][]{\url{#2}}

\bibitem[Aggarwal et al.(2019)]{11yr_cw} Aggarwal, K., Arzoumanian, Z., Baker, P.~T., et al.\ 2019, \apj, 880, 116. doi:\url{https://doi.org/10.3847/1538-4357/ab2236}

\bibitem[Arzoumanian et al.(2015)]{5yr_bwm} Arzoumanian, Z., Brazier, A., Burke-Spolaor, S., et al.\ 2015, \apj, 810, 150. doi:\url{https://doi.org/10.1088/0004-637X/810/2/150}

\bibitem[Arzoumanian et al.(2016)]{9yr_gwb} Arzoumanian, Z., Brazier, A., Burke-Spolaor, S., et al.\ 2016, \apj, 821, 13. doi:\url{https://doi.org/10.3847/0004-637X/821/1/13}

\bibitem[Arzoumanian et al.(2018)]{11yr_gwb} Arzoumanian, Z., Baker, P.~T., Brazier, A., et al.\ 2018, \apj, 859, 47. doi:\url{https://doi.org/10.3847/1538-4357/aabd3b}

\bibitem[Arzoumanian et al.(2020)]{12p5yr_gwb} Arzoumanian, Z., Baker, P.~T., Blumer, H., et al.\ 2020, \apjl, 905, L34. doi:\url{https://doi.org/10.3847/2041-8213/abd401}

\bibitem[Arzoumanian et al.(2021)]{12p5yr_pt} Arzoumanian, Z., Baker, P.~T., Blumer, H., et al.\ 2021, \prl, 127, 251302. doi:\url{https://doi.org/10.1103/PhysRevLett.127.251302}

\bibitem[Bhattacharya et al.(2021)]{Bhattacharya2021} Bhattacharya, S., Mohanty, S., \& Parashari, P.\ 2021, \prd, 103, 063532. doi:\url{https://doi.org/10.1103/PhysRevD.103.063532}

\bibitem[Blanco-Pillado \& Olum(1999)]{BlancoPillado1999} Blanco-Pillado, J.~J. \& Olum, K.~D.\ 1999, \prd, 59, 063508. doi:\url{https://doi.org/10.1103/PhysRevD.59.063508}

\bibitem[Blanco-Pillado \& Olum(2017)]{BlancoPillado2017} Blanco-Pillado, J.~J. \& Olum, K.~D.\ 2017, \prd, 96, 104046. doi:\url{https://doi.org/10.1103/PhysRevD.96.104046}

\bibitem[Blanco-Pillado et al.(2018)]{BlancoPillado2018} Blanco-Pillado, J.~J., Olum, K.~D., \& Siemens, X.\ 2018, Physics Letters B, 778, 392. doi:\url{https://doi.org/10.1016/j.physletb.2018.01.050}

\bibitem[Blasi et al.(2021)]{Blasi2021} Blasi, S., Brdar, V., \& Schmitz, K.\ 2021, \prl, 126, 041305. doi:\url{https://doi.org/10.1103/PhysRevLett.126.041305}

\bibitem[Boileau et al.(2021)]{Boileau2021} Boileau, G., Christensen, N., Meyer, R., et al.\ 2021, \prd, 103, 103529. doi:\url{https://doi.org/10.1103/PhysRevD.103.103529}

\bibitem[Burke-Spolaor et al.(2019)]{BurkeSpolaor2019} Burke-Spolaor, S., Taylor, S.~R., Charisi, M., et al.\ 2019, \aapr, 27, 5. doi:\url{https://doi.org/10.1007/s00159-019-0115-7}

\bibitem[Caprini et al.(2010)]{Caprini2010} Caprini, C., Durrer, R., \& Siemens, X.\ 2010, \prd, 82, 063511. doi:\url{https://doi.org/10.1103/PhysRevD.82.063511}

\bibitem[Carlin \& Chib (1995)]{Carlin1995} Carlin, B. \& Chib, S. Bayesian Model Choice via Markov Chain Monte Carlo Methods. {\em Journal Of The Royal Statistical Society. Series B (Methodological)}. \textbf{57}, 473-484 (1995), \url{http://www.jstor.org/stable/2346151}

\bibitem[Chen et al.(2019)]{Chen2019} Chen, S., Sesana, A., \& Conselice, C.~J.\ 2019, \mnras, 488, 401. doi:\url{https://doi.org/10.1093/mnras/stz1722}

\bibitem[De Luca et al.(2021)]{DeLuca2021} De Luca, V., Franciolini, G., \& Riotto, A.\ 2021, \prl, 126, 041303. doi:\url{https://doi.org/10.1103/PhysRevLett.126.041303}

\bibitem[Deryagin et al.(1986)]{Deryagin1986} Deryagin, D.~V., Grigoriev, D.~Y., Rubakov, V.~A., et al.\ 1986, Modern Physics Letters A, 1, 593. doi:\url{https://doi.org/10.1142/S0217732386000750}

\bibitem[Edwards et al.(2006)]{TEMPO2II} Edwards, R.~T., Hobbs, G.~B., \& Manchester, R.~N.\ 2006, \mnras, 372, 1549. doi:\url{https://doi.org/10.1111/j.1365-2966.2006.10870.x}

\bibitem[Ellis \& Lewicki(2021)]{Ellis2021} Ellis, J. \& Lewicki, M.\ 2021, \prl, 126, 041304. doi:\url{https://doi.org/10.1103/PhysRevLett.126.041304}

\bibitem[Ellis \& van Haasteren (2017]{PTMCMC} Ellis, Justin A., \& van Haasteren, Rutger (2017). PTMCMCSampler (v1.0.0). Zenodo. \url{https://doi.org/10.5281/zenodo.1037579}

\bibitem[Ellis et al.(2020)]{Enterprise} Ellis, Justin A., Vallisneri, Michele, Taylor, Stephen R., \& Baker, Paul T. (2020). ENTERPRISE: Enhanced Numerical Toolbox Enabling a Robust PulsaR Inference SuitE (v3.0.0). Zenodo. \url{https://doi.org/10.5281/zenodo.4059815}

\bibitem[Fabbri \& Pollock(1983)]{Fabbri1983} Fabbri, R. \& Pollock, M.~D.\ 1983, Physics Letters B, 125, 445. doi:\url{https://doi.org/10.1016/0370-2693(83)91322-9}

\bibitem[Godsill (2001)]{Godsill2001} Godsill, S. {\em Journal Of Computational And Graphical Statistics}. \textbf{10}, 230-248 (2001), doi:\url{https://doi.org/10.1198/10618600152627924}

\bibitem[Grishchuk(1976)]{Grishchuk1976} Grishchuk, L.~P.\ 1976, Soviet Journal of Experimental and Theoretical Physics Letters, 23, 293

\bibitem[Grishchuk(1977)]{Grishchuk1977} Grishchuk, L.~P.\ 1977, Eighth Texas Symposium on Relativistic Astrophysics, 302, 439. doi:\url{https://doi.org/10.1111/j.1749-6632.1977.tb37064.x}

\bibitem[Grishchuk(2005)]{Grishchuk2005} Grishchuk, L.~P.\ 2005, Physics Uspekhi, 48, 1235. doi:\url{https://doi.org/10.1070/PU2005v048n12ABEH005795}

\bibitem[Harris et al.(2020)]{Numpy2020} Harris, C.~R., Millman, K.~J., van der Walt, S.~J., et al.\ 2020, \nat, 585, 357. doi:\url{https://doi.org/10.1038/s41586-020-2649-2}

\bibitem[Hazboun et al.(2020)]{Hazboun2020} Hazboun, J.~S., Simon, J., Siemens, X., et al.\ 2020, \apjl, 905, L6. doi:\url{https://doi.org/10.3847/2041-8213/abca92}

\bibitem[Hee et al.(2016)]{Hee2016} Hee, S., Handley, W.~J., Hobson, M.~P., et al.\ 2016, \mnras, 455, 2461. doi:\url{https://doi.org/10.1093/mnras/stv2217}

\bibitem[Hellings \& Downs(1983)]{Hellings1983} Hellings, R.~W. \& Downs, G.~S.\ 1983, \apjl, 265, L39. doi:\url{https://doi.org/10.1086/183954}

\bibitem[Hobbs et al.(2006)]{TEMPO2I} Hobbs, G., Edwards, R., \& Manchester, R.\ 2006, Chinese Journal of Astronomy and Astrophysics Supplement, 6, 189

\bibitem[Hobbs et al.(2009)]{TEMPO2III} Hobbs, G., Jenet, F., Lee, K.~J., et al.\ 2009, \mnras, 394, 1945. doi:\url{https://doi.org/10.1111/j.1365-2966.2009.14391.x}

\bibitem[Hogan(1986)]{Hogan1986} Hogan, C.~J.\ 1986, \mnras, 218, 629. doi:\url{https://doi.org/10.1093/mnras/218.4.629}

\bibitem[Hunter(2007)]{Matplotlib2007} Hunter, J.~D.\ 2007, {\em Comput. Sci. Eng.\/}, {\bf 9}, 90

\bibitem[Jaffe \& Backer(2003)]{Jaffe2003} Jaffe, A.~H. \& Backer, D.~C.\ 2003, \apj, 583, 616. doi:\url{https://doi.org/10.1086/345443}

\bibitem[Kamionkowski \& Kovetz(2016)]{Kamionkowski2016} Kamionkowski, M. \& Kovetz, E.~D.\ 2016, \araa, 54, 227. doi:\url{https://doi.org/10.1146/annurev-astro-081915-023433}

\bibitem[Kelley et al.(2017)]{Kelley2017} Kelley, L.~Z., Blecha, L., \& Hernquist, L.\ 2017, \mnras, 464, 3131. doi:\url{https://doi.org/10.1093/mnras/stw2452}

\bibitem[Kelley et al.(2018)]{Kelley2018} Kelley, L.~Z., Blecha, L., Hernquist, L., et al.\ 2018, \mnras, 477, 964. doi:\url{https://doi.org/10.1093/mnras/sty689}

\bibitem[Kibble(1976)]{Kibble1976} Kibble, T.~W.~B.\ 1976, Journal of Physics A Mathematical General, 9, 1387. doi:\url{https://doi.org/10.1088/0305-4470/9/8/029}

\bibitem[Kobakhidze et al.(2017)]{Kobakhidze2017} Kobakhidze, A., Lagger, C., Manning, A., et al.\ 2017, European Physical Journal C, 77, 570. doi:\url{https://doi.org/10.1140/epjc/s10052-017-5132-y}

\bibitem[Lasky et al.(2016)]{Lasky2016} Lasky, P.~D., Mingarelli, C.~M.~F., Smith, T.~L., et al.\ 2016, Physical Review X, 6, 011035. doi:\url{https://doi.org/10.1103/PhysRevX.6.011035}

\bibitem[Linde(1982)]{Linde1982} Linde, A.~D.\ 1982, Physics Letters B, 108, 389. doi:\url{https://doi.org/10.1016/0370-2693(82)91219-9}

\bibitem[Millman \& Aivazis(2011)]{Python2011} Millman, K.~J. \& Aivazis, M.\ 2011, {\em Comput. Sci. Eng.\/}, {\bf 13}, 9

\bibitem[Mingarelli et al.(2017)]{Mingarelli2017} Mingarelli, C.~M.~F., Lazio, T.~J.~W., Sesana, A., et al.\ 2017, Nature Astronomy, 1, 886. doi:\url{https://doi.org/10.1038/s41550-017-0299-6}

\bibitem[Oliphant(2007)]{Python2007} Oliphant, T.~E.\ 2007, {\em Comput. Sci. Eng.\/}, {\bf 9}, 10

\bibitem[{\"O}lmez et al.(2010)]{Olmez2010} {\"O}lmez, S., Mandic, V., \& Siemens, X.\ 2010, \prd, 81, 104028. doi:\url{https://doi.org/10.1103/PhysRevD.81.104028}

\bibitem[Parida et al.(2016)]{Parida2016} Parida, A., Mitra, S., \& Jhingan, S.\ 2016, \jcap, 2016, 024. doi:\url{https://doi.org/10.1088/1475-7516/2016/04/024}

\bibitem[Parida et al.(2019)]{Parida2019} Parida, A., Suresh, J., Mitra, S., et al.\ 2019, arXiv:1904.05056

\bibitem[Phinney(2001)]{Phinney2001} Phinney, E.~S.\ 2001, astro-ph/0108028

\bibitem[Planck Collaboration et al.(2020)]{Planck2020} Planck Collaboration, Aghanim, N., Akrami, Y., et al.\ 2020, \aap, 641, A6. doi:\url{https://doi.org/10.1051/0004-6361/201833910}

\bibitem[Pol et al.(2021)]{Pol2021} Pol, N.~S., Taylor, S.~R., Zoltan Kelley, L., et al.\ 2021, \apjl, 911, L34. doi:\url{https://doi.org/10.3847/2041-8213/abf2c9}

\bibitem[Rajagopal \& Romani(1995)]{Rajagopal1995} Rajagopal, M. \& Romani, R.~W.\ 1995, \apj, 446, 543. doi:\url{https://doi.org/10.1086/175813}

\bibitem[Romano et al.(2021)]{Romano2021} Romano, J.~D., Hazboun, J.~S., Siemens, X., et al.\ 2021, \prd, 103, 063027. doi:\url{https://doi.org/10.1103/PhysRevD.103.063027}

\bibitem[Rosado et al.(2015)]{Rosado2015} Rosado, P.~A., Sesana, A., \& Gair, J.\ 2015, \mnras, 451, 2417. doi:\url{https://doi.org/10.1093/mnras/stv1098}

\bibitem[Sampson et al.(2015)]{Sampson2015} Sampson, L., Cornish, N.~J., \& McWilliams, S.~T.\ 2015, \prd, 91, 084055. doi:\url{https://doi.org/10.1103/PhysRevD.91.084055}

\bibitem[Sesana et al.(2004)]{Sesana2004} Sesana, A., Haardt, F., Madau, P., et al.\ 2004, \apj, 611, 623. doi:\url{https://doi.org/10.1086/422185}

\bibitem[Sesana et al.(2008)]{Sesana2008} Sesana, A., Vecchio, A., \& Colacino, C.~N.\ 2008, \mnras, 390, 192. doi:\url{https://doi.org/10.1111/j.1365-2966.2008.13682.x}

\bibitem[Siemens et al.(2013)]{Siemens2013} Siemens, X., Ellis, J., Jenet, F., et al.\ 2013, Classical and Quantum Gravity, 30, 224015. doi:\url{https://doi.org/10.1088/0264-9381/30/22/224015}

\bibitem[Siemens et al.(2007)]{Siemens2007} Siemens, X., Mandic, V., \& Creighton, J.\ 2007, \prl, 98, 111101. doi:\url{https://doi.org/10.1103/PhysRevLett.98.111101}

\bibitem[Starobinsky(1980)]{Starobinsky1980} Starobinsky, A.~A.\ 1980, Physics Letters B, 91, 99. doi:\url{https://doi.org/10.1016/0370-2693(80)90670-X}

\bibitem[Taylor et al.(2017)]{Taylor2017} Taylor, S.~R., Simon, J., \& Sampson, L.\ 2017, \prl, 118, 181102. doi:\url{https://doi.org/10.1103/PhysRevLett.118.181102}

\bibitem[Taylor et al.(2020)]{Taylor2020} Taylor, S.~R., van Haasteren, R., \& Sesana, A.\ 2020, \prd, 102, 084039. doi:\url{https://doi.org/10.1103/PhysRevD.102.084039}

\bibitem[Taylor et al.(2021)]{enterpriseextensions} Taylor, S.~R., Baker, P.~T., Hazboun, J.~S., Simon, J.~.J., \& Vigeland, S.~J. 2021, {\em enterprise extensions} \url{https://github.com/nanograv/enterprise_extensions}

\bibitem[Thrane \& Romano(2013)]{Thrane2013} Thrane, E. \& Romano, J.~D.\ 2013, \prd, 88, 124032. doi:\url{https://doi.org/10.1103/PhysRevD.88.124032}

\bibitem[Vallisneri et al.(2021)]{libstempo} Vallisneri, Michele, Ellis, Justin A., Taylor, van Haasteren, R., \& Stephen R. (2021). libstempo (v2.4.3). Github. \url{https://github.com/vallis/libstempo}

\bibitem[Vaskonen \& Veerm{\"a}e(2021)]{Vaskonen2021} Vaskonen, V. \& Veerm{\"a}e, H.\ 2021, \prl, 126, 051303. doi:\url{https://doi.org/10.1103/PhysRevLett.126.051303}

\bibitem[Vehtari et al.(2019)]{Vehtari2019} Vehtari, A., Gelman, A., Simpson, D., et al.\ 2019, arXiv:1903.08008

\bibitem[Vilenkin(1981)]{Vilenkin1981} Vilenkin, A.\ 1981, Physics Letters B, 107, 47. doi:\url{https://doi.org/10.1016/0370-2693(81)91144-8}

\bibitem[Vilenkin(1985)]{Vilenkin1985} Vilenkin, A.\ 1985, \physrep, 121, 263. doi:\url{https://doi.org/10.1016/0370-1573(85)90033-X}

\bibitem[Vilenkin \& Shellard(2000)]{Vilenkin2000} Vilenkin, A. \& Shellard, E.~P.~S.\ 2000, Cosmic Strings and Other Topological Defects, by A. Vilenkin and E. P. S. Shellard, pp. 578. ISBN 0521654769. Cambridge, UK: Cambridge University Press, July 2000., 578

\bibitem[Virtanen et al.(2020)]{Scipy2020} Virtanen, P., Gommers, R., Oliphant, T.~E., et al.\ 2020, {\em Nature Methods\/}, {\bf 17}, 261

\bibitem[Winicour(1973)]{Winicour1973} Winicour, J.\ 1973, \apj, 182, 919. doi:\url{https://doi.org/10.1086/152193}

\bibitem[Zhao(2011)]{Zhao2011} Zhao, W.\ 2011, \prd, 83, 104021. doi:\url{https://doi.org/10.1103/PhysRevD.83.104021}

\end{thebibliography}
